\documentclass[a4paper,fleqn,usenatbib,useAMS]{mnras}

\usepackage{graphicx}	
\usepackage{amsmath}	
\usepackage{amssymb}	
\usepackage{multicol}        
\usepackage{bm}		
\usepackage{pdflscape}	
\usepackage{supertabular}

\usepackage[]{media9}

\newcommand{\msun}{$M_\odot$}

\newcommand{\kms}{km~s$^{-1}$}
\newcommand{\mhi}{$M_{\rm \footnotesize\textsc{H\,i}}$}
\newcommand{\mst}{$M_*$}

\newcommand{\hi}{H{\sc\,i}}
\newcommand{\magsec}{mag/arcsec$^2$}

\defcitealias{leisman17a}{L17}

\let\oldhref\href
\renewcommand{\href}[2]{\oldhref{#1}{\hbox{#2}}}

\usepackage[T1]{fontenc}
\usepackage{ae,aecompl}

\usepackage{newtxtext,newtxmath}

\title[Environment of HUDs]{The environment of \hi-bearing ultra diffuse galaxies in the ALFALFA survey}

\author[Janowiecki, Jones, Leisman, \& Webb]{Steven Janowiecki,$^{1,2}$\thanks{Contact e-mail: \href{mailto:janowiecki@utexas.edu}{janowiecki@utexas.edu}}
Michael G. Jones,$^{3}$
Lukas Leisman,$^{4}$
Andrew Webb$^{4}$
\\
$^{1}$International Centre for Radio Astronomy Research (ICRAR), University of Western Australia, 35 Stirling Highway, Crawley,WA 6009, Australia\\
$^{2}$University of Texas, Hobby-Eberly Telescope, McDonald Observatory, TX 79734, USA\\
$^{3}$Instituto de Astrof\'{i}sica de Andaluc\'{i}a (CSIC), Glorieta de la Astronom\'{i}a, 18008 Granada, Spain\\
$^{4}$Department of Physics and Astronomy, Valparaiso University, Valparaiso, IN 46383, USA
}

\date{ Accepted 2019 June 26. Received 2019 June 6; in original form 2019 January 18}

\pubyear{2019}

\begin{document}
\label{firstpage}
\pagerange{\pageref{firstpage}--\pageref{lastpage}}
\maketitle

\begin{abstract}
We explore the environment of 252 \hi-bearing Ultra Diffuse Galaxies (HUDs) from the 100\% ALFALFA survey catalog in an attempt to constrain their formation mechanism. We select sources from ALFALFA with surface brightnesses, magnitudes, and radii consistent with other samples of Ultra Diffuse Galaxies (UDGs), without restrictions on their isolation or environment, more than doubling the previously reported ALFALFA sample. We quantify the galactic environment of HUDs using several metrics, including n-th nearest neighbour, tidal influence, membership in a group/cluster, and distance from nearest group/cluster or filament.
We find that 
that HUDs inhabit the same environments as other samples of \hi-selected galaxies and that they show no environmental preference in any metric. 
We suggest that these results are consistent with a picture of the extreme properties of HUDs 
being driven by internal mechanisms and that they are largely unperturbed by environmental impacts. While environmental effects may be necessary to convert HUDs into gas-poor cluster UDGs, these effects are not required for diffuse galaxies to exist in the first place.

\end{abstract}




\section{Introduction}

Knowledge of a galaxy's cold gas reservoir is critical to understanding its evolution. 
Recent large neutral hydrogen (\hi) surveys have given direct access to the gas content of tens of thousands of galaxies, providing new insights into questions of ``nature'' vs. ``nurture'': variations in star formation efficiency and the effects of environment on galaxy evolution.

A galaxy's evolution clearly depends in part on its ``nature" - its available fuel supply and rate of consumption. 
For example, \citet{catinella18} have shown that the \hi-to-stellar mass fraction decreases in galaxies of higher stellar masses (\mst), and \citet{huang12b} suggest that \hi-selected galaxies occupy dark matter halos with higher than average spin parameters, suggesting that galaxies with substantial gas reservoirs today may be forming stars less efficiently than their gas-less counterparts did in the past. 

However, external environmental influences (``nurture'') can dramatically affect a galaxy's evolutionary pathway. 
Galaxy morphology correlates with the local density of neighbouring galaxies, \citep[e.g.,][]{dressler80a} as early-type (red, passive) galaxies are more frequently found in high density regions while lower density regions more often host late-type galaxies (blue, star-forming). Studies of the \hi \, mass function have shown that the characteristic \hi \, mass is larger at higher local densities \citep{jones16a}, although the highest local densities observed for optically-selected galaxies are seldom realised for \hi-selected galaxies. 

On larger scales, membership in a group or cluster 
not only encodes the density of nearby galaxies (i.e., clusters are dense) but also connects to a particular location within the large scale structure of the cosmic web.  
Galaxies in groups or clusters are found to have less \hi \, than otherwise similar galaxies in the field \citep{giovanelli85a,solanes02a}, and \hi \, content generally decreases with increasing environmental density for galaxies at  fixed stellar mass \citep{denes14}.  
Further, the mass of a group/cluster can also affect the \hi \, in member galaxies \citep{catinella15, hess13a, janowiecki17}.  
Beyond group membership, some studies have considered a galaxy's particular location within large scale structure. 
For example, \cite{moorman14} found that the \hi \, content of galaxies in ``walls'' (denser structures) was higher than that of those in ``voids'' (low density regions). \citet{kleiner17} found that galaxies within 700~kpc of the spines of the cosmic web have elevated \hi, while \citet{odekon18} showed that \hi \, content decreases with distance from filaments, at fixed local density. 

These results have demonstrated that \hi \, content is clearly tied to both internal and environmental aspects of galaxy formation, however to date there has been no such study specific to \hi-bearing Ultra Diffuse galaxies (HUDs), thus whether or not their extreme properties are linked to their environment is a completely open question from an observational standpoint.

Ultra Diffuse Galaxies (UDGs) are a
population of low surface brightness galaxies (LSBs) that are characterized by their extended ($r_{\textrm{eff}}$=$1.5-4.6$~kpc), and yet faint, diffuse ($\mu_{0,g}>$24 \magsec) appearance \citep{abraham14a,vandokkum15a}. Most samples of UDGs are found in clusters, and seem to fall along the red sequence with early-type optical morphologies. Yet this is in part due to the the difficulty of spectroscopic observations for such faint objects: their distances are typically inferred from their (apparent) proximity to a group or cluster with known distance. Follow-up observations have shown these assumptions to be generally accurate, but sometimes UDGs can be members of more distant background structures \citep[e.g.,][]{danieli17}.
Even given these challenges, recent work has identified hundreds of UDGs in new and archival observations across all environments: galaxy clusters \citep{mihos15a,wittmann17}, groups \citep{merritt16a, castelli16a, shi17}, filaments \citep{martinez-delgado16a}, and the field \citep{papastergis17a,leisman17a}.   

Three main formation pathways for UDGs have been discussed in the literature: first, that UDGs are ``failed'' L$^*$ galaxies that lost their gas at early times 
\citep[e.g.,][]{beasley16b, peng16, vandokkum16a}; 
second, that UDGs occupy the high spin tail of the galaxy angular momentum distribution \citep[e.g.,][]{amorisco16a};
third, that UDGs are dwarf galaxies that have been made diffuse by stellar feedback and outflows
\citep[e.g.,][]{collins13}.  
More recent observations and simulations have increasingly favored the latter two explanations for the majority of the UDG population \citep{beasley16a, beasley16b, dicintio17a, papastergis17a, amorisco18a, chan18, ferremateu18a}, suggesting that feedback and internal processes are important to UDG formation; 
however, \citet{bennet18} suggests that some UDGs may have a tidal origin 
and the high-resolution simulations \citet{jiang18a,jiang18b} find that UDGs live in halos with ordinary spin parameters.  

Despite the importance of gas reservoirs to galaxy evolution, very little is known about their presence or role in UDGs. Searches for 21cm emission have focused on UDGs in isolation, where \hi \, is more likely to persist \citep{bellazzini15a,leisman17a,papastergis17a}. Recent work has also detected \hi \, in optically blue UDGs that are a members of Hickson Compact Groups \citep{spekkens18} and a relatively poor galaxy cluster \citep{shi17}. However, a deep search for \hi \, from NGC1052-DF2 yielded a non-detection \citep{chowdhury18}.
Although few HUDs have been imaged with interferometers (to date), limiting our knowledge of the internal gas kinematics, \citet{jones18c} performed a study to assess their abundance and global gas kinematics, relative to the rest of the \hi-selected galaxy population. They found that although far fewer HUDs are known than red UDGs in clusters or groups, their overall cosmic abundances appear to be similar, and that on average HUDs display a strong preference for low velocity widths.
When \hi \, is detected in UDGs it not only informs the evolutionary potential and depletion time \citep[HUDs have very low star formation efficiencies,][]{leisman17a}, but it also provides a redshift without requiring any assumption about cluster membership. This independent distance estimate allows us to quantify the absolute sizes and masses of HUDs and connect them with the population of cluster UDGs.

Recently, \citet[][hereafter \citetalias{leisman17a}]{leisman17a} examined the \hi \, properties of a large sample of HUDs. Using the Arecibo Legacy Fast ALFA (Arecibo L-band Feed Array) extragalactic \hi \, survey \citep{giovanelli05a,haynes11a,haynes18}, \citetalias{leisman17a} identified a sample of isolated HUDs with low surface brightness optical counterparts and no nearby neighbors. 
\citetalias{leisman17a} found that these HUDs were optically bluer and more slowly rotating than typical UDGs, and suggested that they may be a gas-rich progenitor population to UDGs found in clusters.

To extend the work of \citetalias{leisman17a}, we select HUDs without restrictions on their environments or isolation. We compare the environmental preferences of our sample of HUDs with other samples of \hi-detected galaxies to test evolutionary pathways of UDGs. 
If external effects contribute to the diffuse nature of HUDs then we would expect to see an environmental separation between them and the general HI-bearing galaxy population. In addition, if HUDs are a progenitor population of red UDGs, and (rapidly) undergo the transformation when they fall into clusters, then we would not expect to find HUDs in high density environments.

In this work we analyze the environment of HUDs. 
In Section~\ref{sec:sample} we discuss the selection and properties of the HUD sample, and in Section~\ref{sec:methods} we discuss our control sample and the environmental measures we use. In Section~\ref{sec:results} we present our results, and we discuss their implications in Section~\ref{sec:discussion}. We conclude in Section~\ref{sec:conclusion}.

\section{HUDs sample selection criteria}
\label{sec:sample}

We select a sample of HUDs from the full ALFALFA survey, expanding on and improving the sample from  \citetalias{leisman17a}. 
Following \citetalias{leisman17a}, we use "\hi-bearing" to refer to ultra-diffuse galaxies which have enough \hi \, to be detected at the sensitivity limits of the ALFALFA survey within our distance range.   
There are 31,506 high significance extragalactic detections in the 100\% ALFALFA catalog \citep{haynes18}. Following \citetalias{leisman17a}, we apply a minimum distance cut of 25~Mpc to eliminate more nearby sources with significant uncertainties in redshift-dependent distance estimates, and a maximum distance cut of 120~Mpc since optical identification can become uncertain due to Arecibo's increasingly large beam (3.5\arcmin). 
These restrictions also aid in providing uniform environment estimates, as discussed below.
Applying these distance limits leaves 15,054 ALFALFA sources. 

Since optical parameters of low surface brightness galaxies are difficult to quantify automatically with shallow survey photometry, we instead use the Sloan Digital Sky Survey (SDSS) catalog measurements to eliminate the well-measured, high surface brightness sources from our sample. We define a source to be well-measured and with high surface brightness if it has a radius larger than 1.5\arcsec \, (it must be resolved by SDSS), is brighter than 22.5~mag in the $r$ filter, and has a ``high'' average surface brightness in $g$,$r$, and $i$ filters using both exponential and Petrosian radii and magnitudes. Eliminating the 11,440 sources that satisfy all of these criteria leaves 3614 
candidate HUDs that require further visual examination. We note that it is possible that a few of these 11,440 ``high'' surface brightness sources are, in fact, HUDs where SDSS has failed to properly measure their parameters. However, the inclusion of minimum radii and magnitude parameters means that these instances should be quite rare. 

Of the remaining sources, 73 lie within 10\arcmin\ of a star in the Yale Bright Stars Catalog, and an additional 713 of the sources lie outside of the SDSS footprint (as determined from visual inspection), and were removed from the sample to maintain sample uniformity. 
We also choose to remove the 168 remaining sources with no identified optical counterpart, since while a few of these sources may be ultra diffuse galaxies lurking below the SDSS detection limit, \cite{leisman17thesis} demonstrate that almost all of these are \hi \, tidal debris. 

We then visually examine the remaining ALFALFA detections, the vast majority of which are high surface brightness galaxies with some issue affecting the photometric estimates of their radius or flux. Some, however, are truly low surface brightness sources, both well-measured and poorly-measured by SDSS. This visual inspection results in a sample of 425 potential low surface brightness, ``ultra diffuse'' sources.
While deeper imaging surveys are becoming available \citep[e.g.,][]{chambers16,Dey19}, developing analogous extraction and surface brightness profile fitting tools for ALFALFA sources in those datasets is complex and the subject of other ongoing work (\citealt{greco18, greco19}, Greco, in preparation).  

We then fit these sources using in-house photometry on downloaded SDSS images with the same procedure described in \citetalias{leisman17a}. In brief, we chose the galaxy center to be the centroid of the extended optical flux, as determined visually (though for sources with clumpy morphologies and significant evidence of star formation, this may not be the location of the peak flux). We then create surface brightness profiles using simple circular apertures, since inclinations are poorly constrained, and then fit exponential functions to these profiles. We correct for Galactic extinction and the effects of the PSF, but do not correct for the small cosmological surface brightness dimming, consistent with other local Universe studies. We note that while most authors fit Sersic profiles with $n$ free, due to the low S/N of SDSS images at these surface brightnesses, we have forced our fits to have exponential ($n$=1) profiles, in keeping with the average value found for UDGs 
 and typical \hi-rich galaxies. 
By their nature the optical properties of UDGs are difficult to accurately measure, especially with the relatively shallow imaging from SDSS. Deeper follow-up observations will be necessary to fit more accurate surface brightness profiles with more free parameters. In this work we aim to select a sample of UDGs that are meaningfully diffuse in optical images, not to characterize their surface brightness profiles in detail.

Using these measurements we then select sources that match the surface brightness, magnitude, and radius properties of other observationally defined samples of UDGs. Specifically, we choose to follow \citetalias{leisman17a} in defining a  
restrictively selected sample (HUDs-R) of 71 \hi-bearing ultra diffuse sources with half light radii r$_{g,{\rm eff}}>$1.5~kpc, $\mu_{g,0}>24$~mag~arcsec$^{-2}$, and M$_g>-$16.8~mag, and a more broadly selected sample (HUDs-B) of 252 sources with r$_{r,{\rm eff}}>1.5$~kpc, $\left< \mu(r,r_{\rm eff}) \right> >24$~mag~arcsec$^{-2}$, and M$_r>-$17.6 (corresponding to the surface brightness and radius limits from \citealp{vandokkum15a} and \citealp{vanderburg16a} respectively, with absolute magnitude limits as defined by \citetalias{leisman17a}).  
Table~\ref{tab:optical} in Appendix~\ref{sec:optical} gives the observed quantities and optical properties of our sample (available online-only).   

This sample represents an increase of more than 2x the number of ALFALFA HUDs found in \citetalias{leisman17a} for three reasons. First, this sample is selected from the full ALFALFA 100\% catalog, rather than the 70\% catalog, contributing an additional 64 
sources to the sample. Second, we are more conservative in our removal of high surface brightness sources with SDSS photometry, allowing the inclusion of 13 sources that were missed by the \citetalias{leisman17a} cuts. Third, the remaining 60 additional sources are included because we do not include an environmental isolation criteria in our sample selection.  \citetalias{leisman17a} required that their sources could not have a neighbour with a measured redshift within 500~\kms\ and 300~kpc. The removal of this criteria is essential to our goal of studying the environmental properties of HUDs, and it further provides a broader and more robust catalog of all ALFALFA HUDs. As discussed in Section~\ref{sec:sample_properties}, the overall properties of the two samples are quite similar.

\section{Methods for Quantifying Environment}
\label{sec:methods}
Quantifying the environmental dependence of galaxy properties typically involves two galaxy samples: the target galaxy sample in question and a reference sample that is used to define the environment of the galaxies in the first sample.  
Such measurements are most robust when both samples are cut in order to be volume limited - when neither the definition of environment, nor the observable (integrated) properties of the target sample, exhibit a distance dependence due to observational effects.

If the reference sample were not volume-limited then the apparent galaxy number density would depend on distance, as the sensitivity to physical quantities (e.g., luminosity) would change with distance. Therefore, any environment metric that was calculated from this sample without weighting based on prior knowledge of the underlying galaxy population (e.g., the luminosity function) would be biased. Thus, in the absence of such weighting, a non-biased environment metric requires the reference sample to be volume limited.

The target sample also has similar complications because large-scale structure (LSS) varies as a function of distance. If the target sample is not volume limited then an apparent trend with environment could actually be due to a particular type of source being more or less detectable at a given distance that, by chance, coincides with a strong feature in LSS. 

\begin{figure*}
    \centering
    \includegraphics[width=\columnwidth]{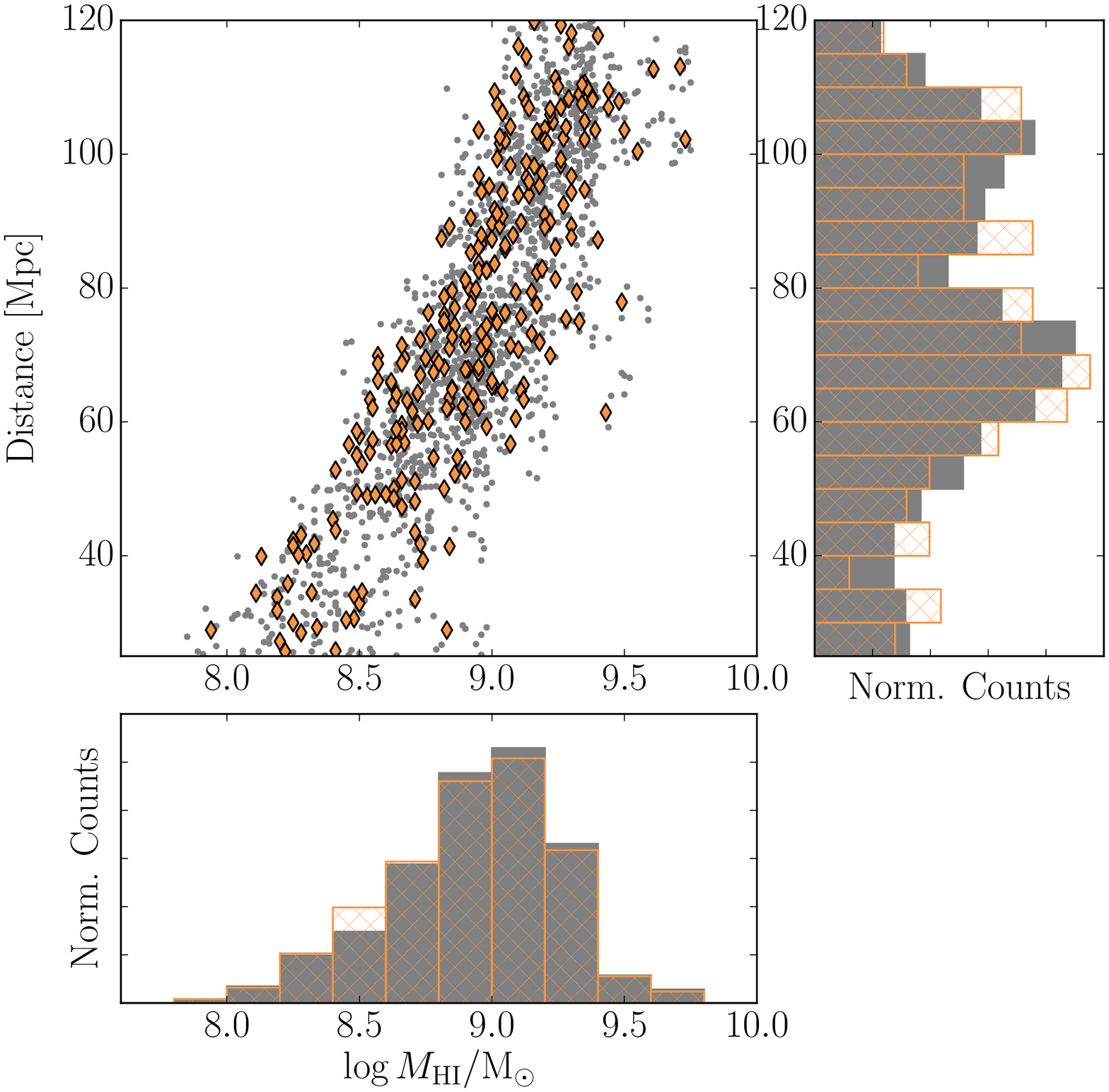}
    \includegraphics[width=\columnwidth]{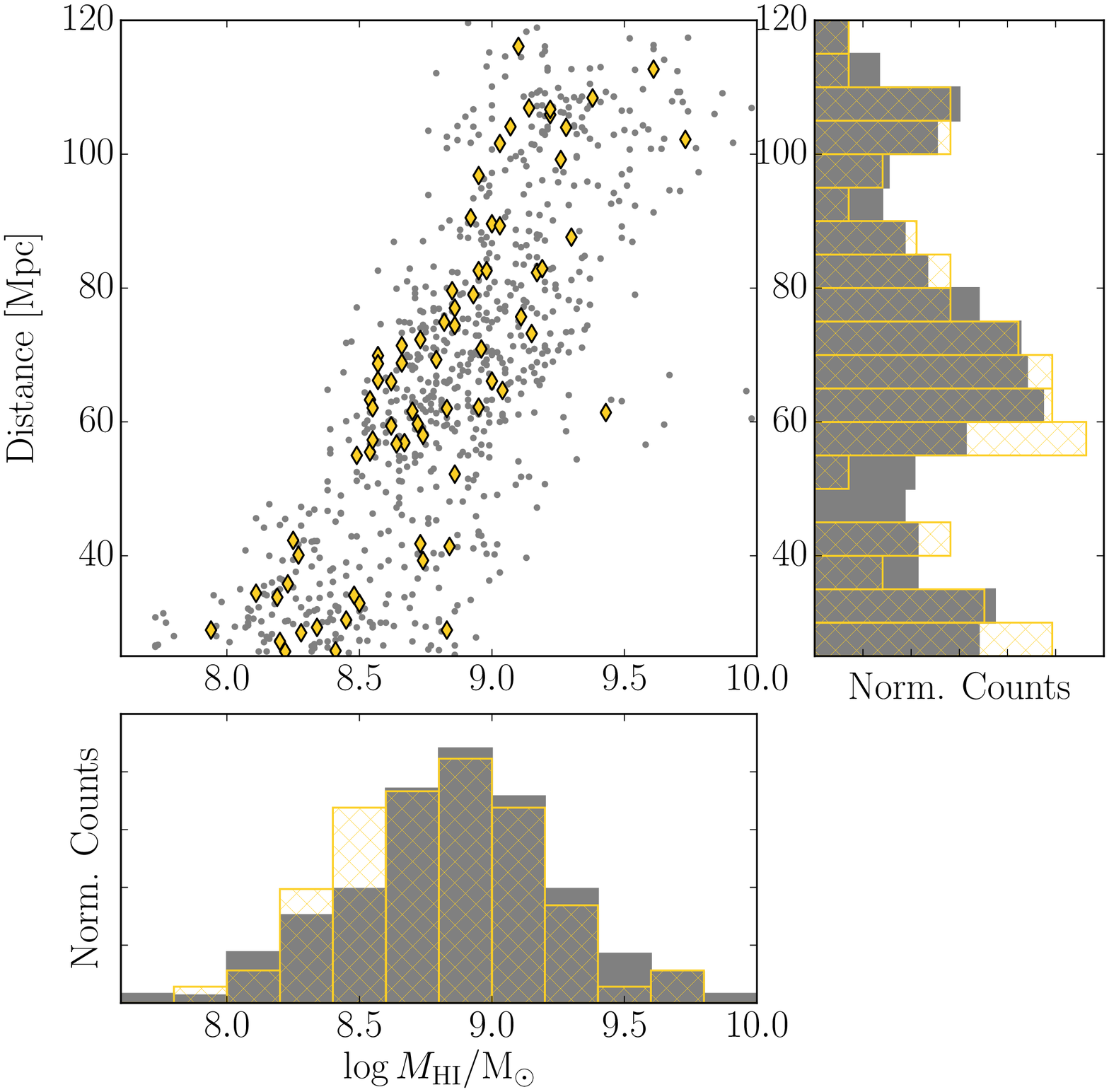}
    \caption{\textit{Left}: The distribution of the \hi \ masses and distances of the HUDs-B target sample (orange diamonds and cross-hatched bins) compared to the matched comparison sample (grey dots and filled bins). \textit{Right}: As for the left figure, but for the HUDs-R target sample.}
    \label{fig:comp_sample}
\end{figure*}

Thus, to appropriately quantify the environment of HUDs we remove distance dependence for both our reference and target samples.
For the reference samples we take the simple approach of cutting them to be volume limited (with the exception of the methods discussed in Sections~\ref{sec:fil_dists} and \ref{sec:grp_dists}, which address volume completeness independently). 
However, we do not wish to incur the reduction in sample size of the target galaxies that would result if we were to cut the sample to be volume limited. Therefore, we choose to use the full target sample but to draw a comparison sample from the rest of the ALFALFA population that is matched to have the same distribution of both distance and \hi \ mass as the target sample. By restricting our environmental comparisons to be between the target sample and the matched comparison sample we eliminate the chance that any potential trends are really a result of the distribution of our sources along the line of sight, or that a correlation between galaxy \hi \ mass and environment is masquerading as a trend related to surface brightness. The matched samples for the HUDs-R and HUDs-B target samples (see Section~\ref{sec:sample}) are shown in Figure~\ref{fig:comp_sample}.
As will be discussed in detail later, the galaxies in this \hi-matched comparison sample are, by construction, typically more luminous than our HUDs. While this also implies that they may have larger stellar masses than HUDs, when considering total baryonic mass the matched comparison sample galaxies are likely to be more similar to the HUDs.

In the following subsections we first describe the reference samples that are used to measure environment and then describe the various environment metrics used throughout the rest of this paper to quantify the environment of our target sample.

\subsection{Reference samples}
\label{sec:ref_sample}

We make use of five different reference catalogues to define our environment metric:
\begin{itemize}
    \item \textit{SDSS DR13 galaxy sample (spectroscopic)} -- All primary spectroscopic objects classified as galaxies, with clean photometry, and falling in the redshift range 1250 \kms \ to 8900 \kms \ (extended 500 \kms \ beyond the HUD sample redshift range) are selected from DR13 \citep{SDSSDR13} using the CasJobs server. The sample is then cut to be volume limited over this range by setting the maximum r-band Petrosian absolute magnitude\footnote{Distances were calculated using ALFALFA's flow model \citep{masters05b}.} of $M_{\mathrm{r}} < -17.77$. This corresponds to an apparent magnitude of 17.75 at the outer distance limit. We choose to use 17.75 to be slightly more conservative than the nominal value of 17.77, below which the spectroscopic sample is designed to be complete. Furthermore, we include only the portion of the catalogue that was in the Northern Spring sky, as the legacy spectroscopy is highly incomplete in the Fall.
    
    \item \textit{SDSS DR13 galaxy sample (photometric)} -- All primary photometric objects classified as galaxies, with clean photometry, and r-band Petrosian magnitudes brighter than 19, are selected from DR13 using the CasJobs server. To remove excess stars and distant background galaxies we further require that $0 < z_{\mathrm{photo}} < 0.04$ and that the Petrosian radius is greater than $2''$.
    
    \item \textit{2MASS Redshift Survey (2MRS)} -- All galaxies in the redshift range 1250 \kms \ to 8900 \kms \ are selected from the 2MRS \citep{huchra12a} catalogue. The sample is cut to be volume limited by imposing a maximum K-band absolute magnitude of $M_{\mathrm{K}} < -23.77$, corresponding to the limiting apparent magnitude of the survey (11.75) at the outer edge of the redshift range.
    
    \item \textit{Galaxy filament catalog of \citet{tempel14a}} -- The galaxies used in this catalog come from the main contiguous area of the SDSS DR8 \citep{aihara11a} spectroscopic sample, have $r$-band Petrosian magnitudes brighter than $17.77$ \citep{strauss02}, and have redshifts between $0.009 < z < 0.155$. Within this sample volume, a filamentary network is traced in the galaxy distribution using a Bisous model \citep{tempel14a}.

    \item \textit{Galaxy group catalog of \citet{yang07a}} -- This group catalog is an updated version\footnote{obtained from http://gax.shao.ac.cn/data/Group.html} using the SDSS DR7 \citep{abazajian09} spectroscopic sample, but originally used the SDSS DR4 \citep{adelman-mccarthy06a} sample between $0.01 \le z \le 0.20$ and other available redshift catalogs \citep{colless01,saunders00a,rc3}. Using a halo-based friends-of-friends group finder, \citet{yang07a} identified groups of galaxies in this sample based on their redshift and position on the sky.

\end{itemize}

\subsection{Nearest neighbour density}
\label{sec:metric_nnden}

One of the most commonly used environment metrics is neighbour density \citep[e.g.,][and references therein]{baldry06,brough11,muldrew12}, which is a measure of the density of neighbours in the immediate vicinity of a given galaxy. This is typically a local measure of environment, probing within a few Mpc of the target. However, by using a reference catalogue that is sparser and that primarily contains larger red galaxies (2MRS), we also use this method to test somewhat larger scales.

For this metric we use the second nearest neighbour density:
\begin{equation}
\Sigma_{2} = \frac{2}{\pi r^{2}},
\end{equation}
where $r$ is the projected separation to the 2nd nearest neighbour within $\pm$500 \kms. There is also an exclusion zone of 5'' and 70 \kms \ around each galaxy in the HUDs sample, within which neighbours are ignored as they are assumed to be the target galaxy. Although choosing a more distant neighbour would reduce the noise in this metric, we choose only the second neighbour for two reasons: a) due to the volume limit cut the reference sample is much more sparsely populated than it would otherwise be, so in many cases the second neighbour that we identify will actually be a more distant neighbour; and b) because the farther out from the central object we search, the larger overlap with SDSS we require, to not produce bias values for nearby objects (see below).

When using the SDSS spectroscopic reference catalogue the edges of the ALFALFA footprint must be trimmed to ensure more than complete overlap with SDSS. The edges are clipped such that there is always at least 10$^\circ$ of overlap. At a distance of 25 Mpc (the inner edge of the HUDs sample) this corresponds to an overlap of 4.36 Mpc. Thus, neighbour density measurements with $\log \Sigma_{2}/\mathrm{Mpc^{-2}} > -1.47$ will not be influenced by the survey boundary.

\begin{figure}
    \centering
    \includegraphics[width=\columnwidth]{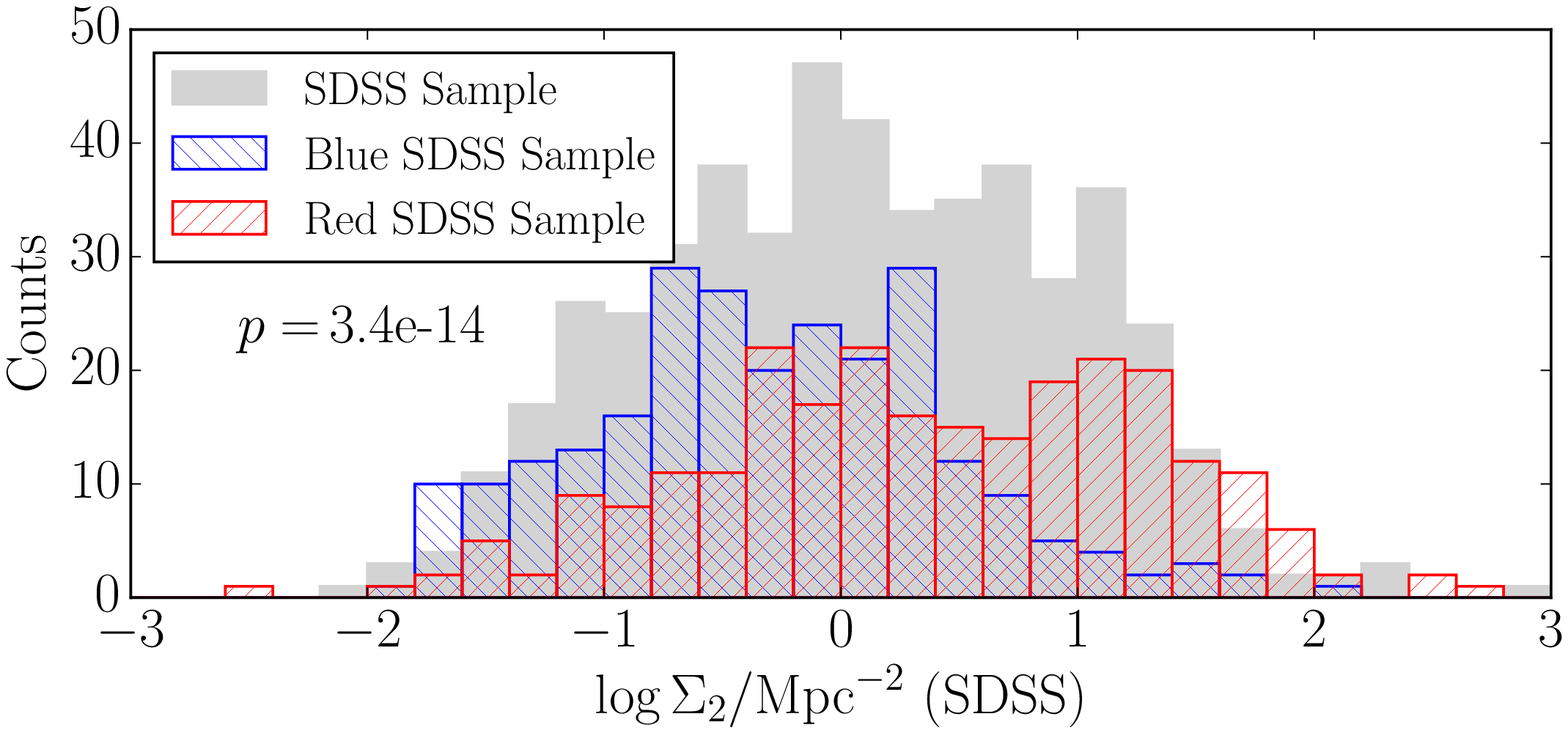}
    \includegraphics[width=\columnwidth]{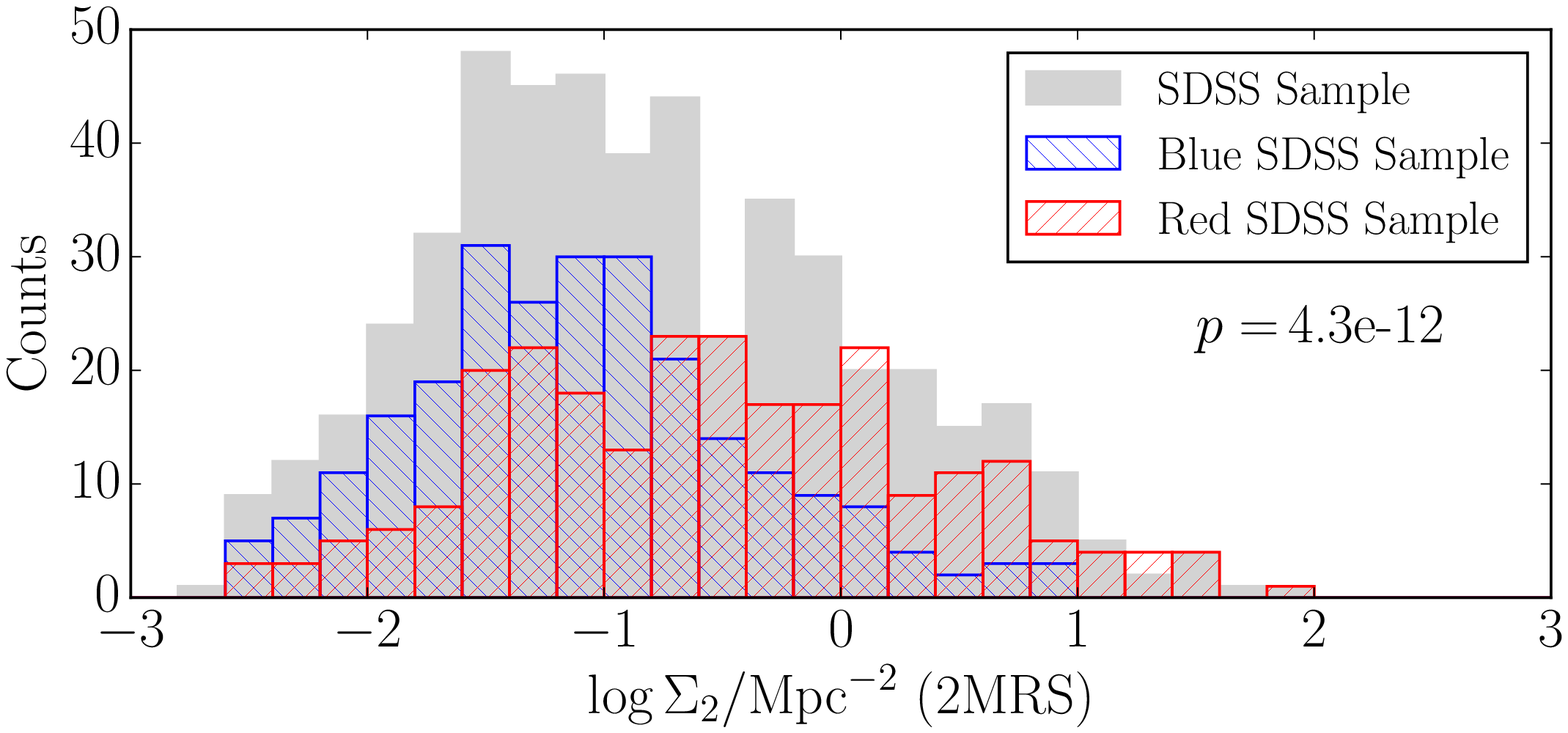}
    \includegraphics[width=\columnwidth]{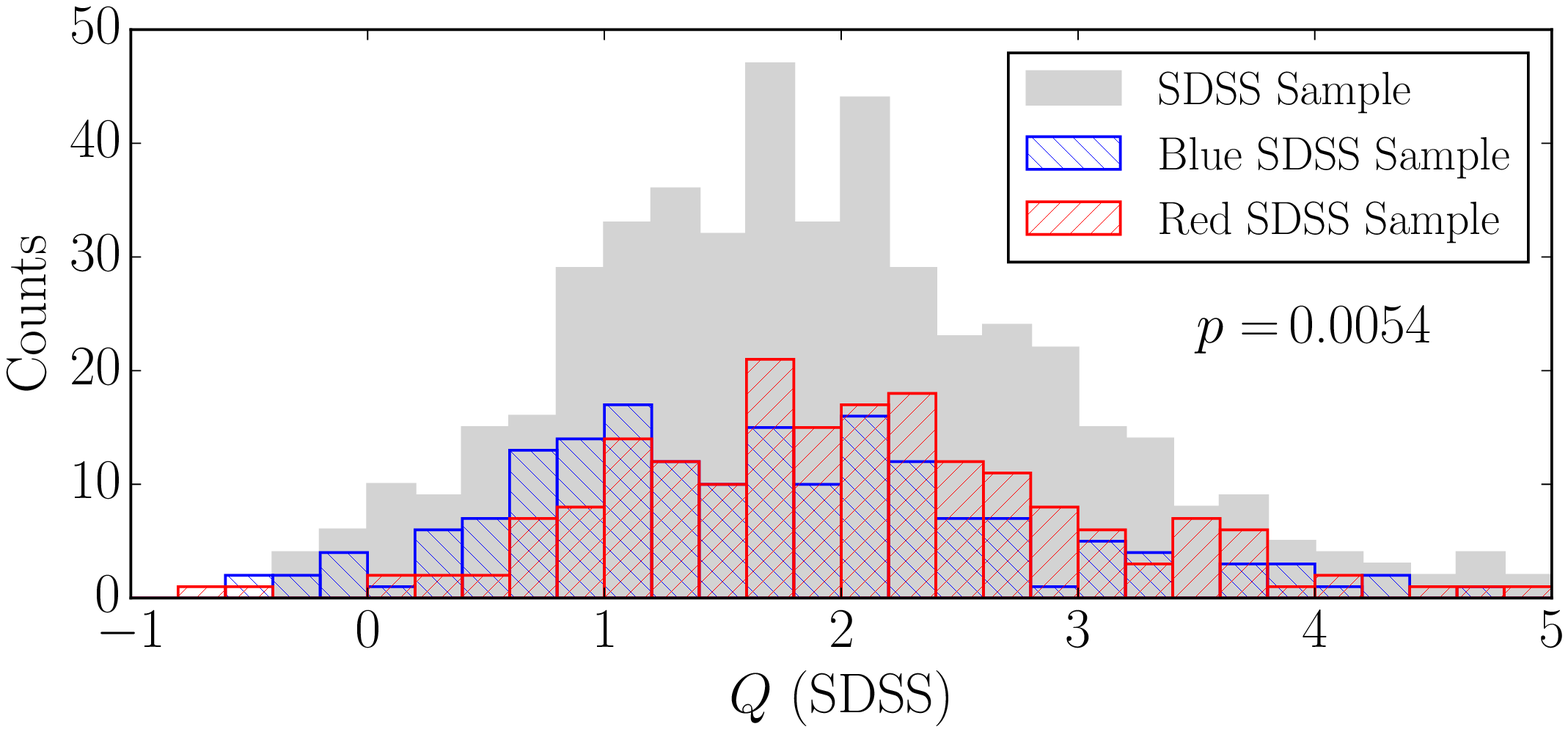}
    \caption{
        \textit{Top}: The distribution of neighbour densities (measured by the SDSS spectroscopic reference catalogue) for randomly selected samples of 250 red ($g-i > 0.85$), 250 blue ($g-i < 0.85$), and 500 SDSS galaxies in the range $25 < D/\mathrm{Mpc} < 120$. 
        \textit{Middle}: As above but with neighbour densities measured by the 2MRS reference catalogue.
        \textit{Bottom}: As above but with tidal strength parameter measured using the SDSS photometric reference catalogue.
        In all cases, the KS test carried out with the red and blue distributions returns a p-value (that the distributions are identical) of significantly below 0.01.
        }
    \label{fig:sdsscomp_ALL}
\end{figure}

The SDSS volume-limited reference catalogue contains 8075 galaxies that fall within the same area and distance range as the HUDs. To demonstrate the effectiveness of these environmental indicators to identify galaxy populations,
this sample is divided approximately in half with a colour cut at $g-i = 0.85$, which roughly separates the red and blue peaks in the bimodal colour distribution. $\Sigma_2$ is calculated for each of these galaxies in an identical manner as for the HUDs. Comparing the $\Sigma_2$ distributions of the red and the blue SDSS galaxies with the KS test gives a KS-statistic of 0.312 (or 0.253 when using $\Sigma_2$ calculated with 2MRS) and a corresponding p-value of $5.94 \times 10^{-172}$, clearly demonstrating that on average red and blue galaxies reside in different environments. However, how the KS-statistic relates to the p-value depends on the effective sample size, $\mu = nm/(n+m)$, where $n$ and $m$ are the sizes of the two samples. Given the separation between the two distributions (i.e., the KS-statistic, 0.312), this separation would be expected to be detectable at a 2-$\sigma$ level ($p < 0.05$) for $\mu > 18.9$ (or 28.8 for 2MRS). In other words, we would be able to detect an environmental shift between the HUDs population and the general \hi \ population, of equivalent scale to that between the SDSS red and blue populations, with a sample of only $\sim$20 ($\sim$30 for 2MRS) HUDs (as $\mu \approx m$ if $n \gg m$). The top two panels of Figure~\ref{fig:sdsscomp_ALL} show the distribution of $\Sigma_2$ for a random sample of 250 red and 250 blue SDSS galaxies to illustrate the scale of this difference for a sample similar in size to the HUDs sample.

\subsection{Tidal influence}
\label{sec:metric_tidal}

This metric estimates the strength of the tidal forces on a galaxy due to its neighbours. One of its advantages is that it can be calculated using only photometric data, which permits more sky to be covered (increasing the sample size), and it incorporates fainter neighbours that may not be included in the spectroscopic sample. However, the drawback of using a photometric metric is that it is much more prone to the effects of interlopers that appear close to the target but that are actually foreground or background sources. We have attempted to mitigate this as much as possible in the reference catalogue selection (Section \ref{sec:ref_sample}), but this metric will still be considerably more noisy than neighbour densities based on the spectroscopic catalogues.

\citet{dahari84} used the optical diameter of a galaxy to the power of 1.5 as a proxy for its mass and created a dimensionless metric, $q_{\mathrm{p}i}$, for the tidal influence of neighbours given by:
\begin{equation}
    q_{\mathrm{p}i} = \frac{(D_{\mathrm{p}}D_{i})^{1.5}}{S_{\mathrm{p}i}^{3}},
\end{equation}
where $D$ is the diameter of a galaxy, $\mathrm{p}$ denotes the target galaxy, $i$ the $i$th neighbour, and $S_{\mathrm{p}i}$ their separation in the same units as $D$. This metric was used by \citet{verley07} to help characterize the AMIGA \citep[Analysis of the interstellar Medium of Isolated Galaxies,][]{verdes05} sample of isolated galaxies, and was updated by \citet{argudo14} who instead used $r$-band luminosities as the proxy for mass. 
We adopt this latter approach, but since it is unclear whether HUDs fall on standard scaling relations, and because their current photometry is poor, we choose not to use the $r$-filter 
luminosity of the target HUD in this calculation. As setting reference $r$-filter luminosity values   
for use with all targets would just amount to setting a zero-point for $Q$, we simply ignore these altogether.  
However, it is important to note that this choice means that the absolute numerical values of $Q$ calculated here cannot be compared to other works. This gives the final expression for $Q$, given by:
\begin{equation}
    Q = \log \sum_{i} q_{\mathrm{p}i} = \log \sum_{i} \frac{10^{-0.4m_{r}^{i}}}{S_{\mathrm{p}i}^{3}},
\end{equation}
where $m_{r}^{i}$ is the apparent $r$-band magnitude of the $i$th neighbour. For the full calculation of $Q$ we consider all photometric neighbours within the catalogue that fall within a projected distance of 1 Mpc of the target galaxy.

The bottom panel of Figure~\ref{fig:sdsscomp_ALL} shows a similar comparison to those discussed above, but now between the $Q$ values of 250 red and blue SDSS galaxies (randomly selected). Although this metric appears less sensitive to the different environments of the two populations, as is to be expected given its purely photometric definition, the null hypothesis that these 500 galaxies are all from the same environment is still rejected with over 3-$\sigma$ confidence. In this case the KS-statistic is 0.152 for the full red and blue samples from SDSS (each $\sim$4000 objects), which would require an effective sample size of $\mu > 79.8$ in order to be detectable at 2-$\sigma$ confidence. Therefore, due to the increased noise in this method, at least $\sim$80 sources would be required to detect the difference between the SDSS red and blue populations.

\begin{figure*}
    \centering
    \includegraphics[width=0.75\textwidth]{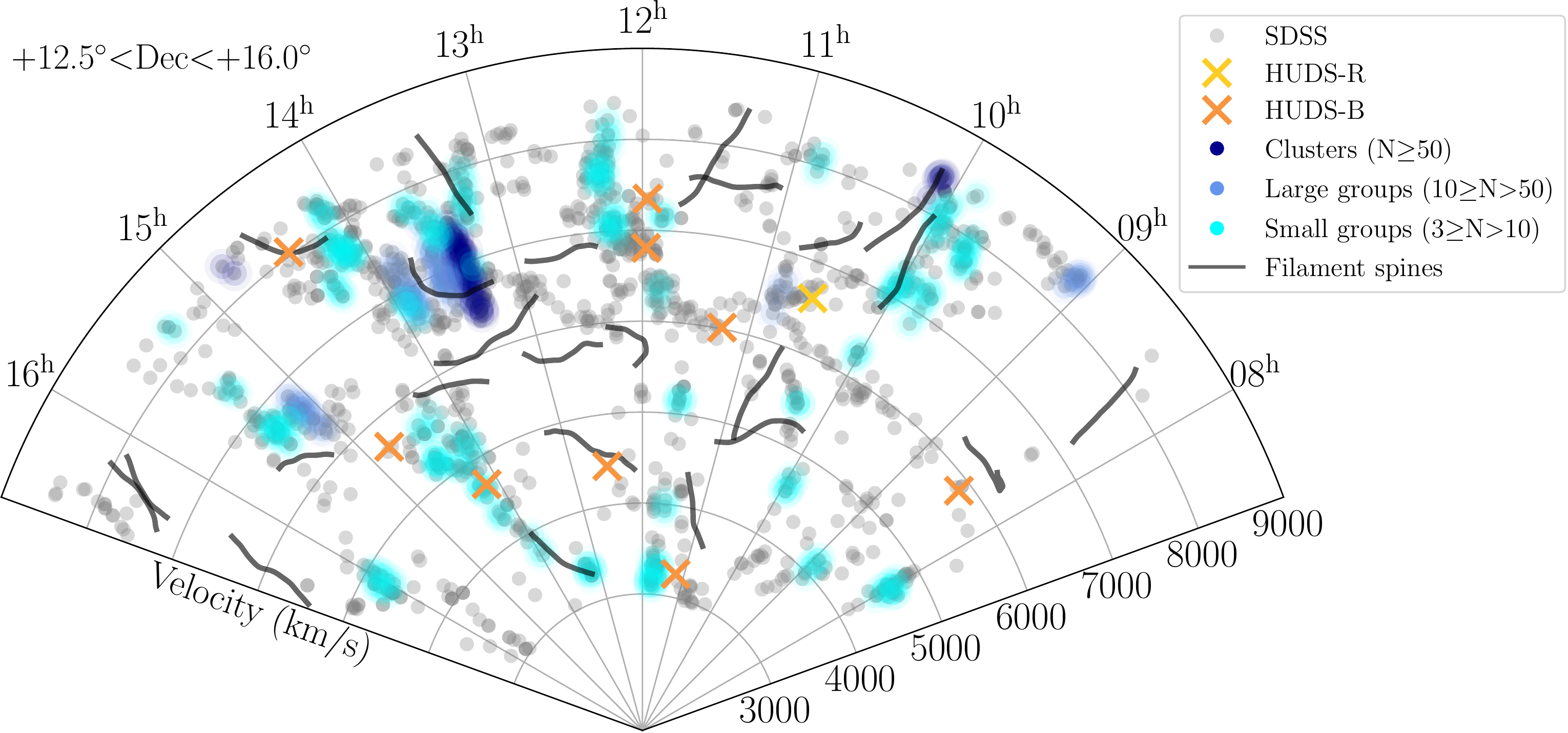}
    \caption{Polar plot of HUDs samples with comparison samples, in a slice of declination between +12.5$^\circ$ and +16$^\circ$ in the Northern Spring sky. Clusters/groups from the \citet{yang07a} catalog and filament spines from \citet{tempel14a} are also labeled. Note that the HUDs samples largely avoid dense environments.}
    \label{fig:cone}
\end{figure*}

\subsection{Distance from nearest filament}
\label{sec:fil_dists}

To quantify the location of a galaxy within the cosmic web of large scale structure, we determine its distance to the nearest filament spine. Similarly to the previously discussed metrics, volume-limited reference catalogs are used to define the environments of galaxies in our sample. We adopt the filament catalog of \citet{tempel14a}, which used the contiguous (Northern Spring) sky area from SDSS DR8 \citep{aihara11a}. \cite{tempel14a} used the Bisous model to identify filamentary structures in the three-dimensional galaxy distribution. They worked in a probabilistic Bayesian framework and adopted a filament radius of $0.5$$h^{-1}$~Mpc. Previous work using these catalogs has shown that being inside a filament affects galaxy properties even when comparing with non-filament galaxies at fixed local (aperture) density \citep{poudel17a}.

When determining the distance to the nearest filament for each member of our HUDs samples, we first limit our calculations to those that are fully enclosed by the volume of the \citet{tempel14a} catalog. This retains 34/71 of the HUDs-R, 105/183 of the HUDs-B samples, and $\sim$60\% of the matched comparison samples. 
For each included member of our HUDs (and matched comparison) samples we calculate the three-dimensional distance (using sky coordinates plus redshift) to the nearest filament spine in co-moving coordinates. If a galaxy is less than 0.5~Mpc from a filament spine it would be considered ``inside'' that filament.

We find that while only $\sim$1\% of HUDs are located inside filaments (i.e.,
$d_\textrm{filament}$$\le$0.5~Mpc), $\sim$70\% are found within 5~Mpc of a filament spine. The matched comparison sample galaxies show the same occupation fractions. Further analysis of filament occupancy and the  effects of being in or nearby a filament are considered in Section~\ref{sec:results}.

\subsection{Membership and distance from nearest groups}
\label{sec:grp_dists}

To explore the environment in terms of galaxy groups and clusters, we adopt the \citet{yang07a} group catalog, which was generated by a friends-of-friends algorithm to identify groups/clusters within SDSS DR7 \citep{abazajian09}. We use version ``B'', which supplemented the SDSS redshifts with those from other spectroscopic surveys as well. For each galaxy, the catalog identifies whether it is a central galaxy in a group (i.e., most massive member of its group/cluster), a central galaxy in isolation (i.e., the only galaxy in its dark matter halo), or a satellite member of a group/cluster. The group catalog also includes an estimate of the total dark matter halo mass using abundance matching.

As with our comparison to the filament catalog, we consider only the HUDs and matched comparison galaxies that fall inside the volume probed by the \citet{yang07a} group catalog. This reduces our sample to 36/71 of HUDs-R, 113/183 of HUDs-B, and $\sim$60\% of the matched comparison samples.

We first use this group catalog to determine whether HUDs are members of any existing group. As HUDs are too faint for SDSS spectroscopy, they are not included in the \citet{yang07a} catalog. Instead, we use their \hi \, redshifts to check whether they lie inside the virial radius of any existing group. We find only 7 (10) are within 1 (2) virial radius of any group, suggesting that these 5\% (7\%) are very (somewhat) likely to be members of those groups. Approximately 16\% (30\%) of the comparison sample galaxies are found within 1 (2) virial radius of an existing group. The increased membership fraction of the comparison sample may be partially a result of the larger median optical luminosity (and therefore \mst) of comparison sample galaxies compared with our HUDs samples.  
Larger studies have shown that more luminous galaxies are more spatially correlated and more likely to be found in groups \citep[e.g.,][]{zehavi11}.

Next, we consider the other remaining HUDs and galaxies in the matched comparison samples that are not likely to be members of an existing group, and we determine the distances to their nearest groups. We compute both the projected 2-dimensional distance to the nearest group (adopting the group distance to determine a physical separation), and the 3-dimensional co-moving distance. 
These results are shown visually in Figure~\ref{fig:cone}, which includes the SDSS sample, the groups identified by \citet{yang07a}, and our HUDs and comparison samples. Note that the HUDs are typically not found in groups or clusters.

\section{Results}
\label{sec:results}

Here we present the properties of the extended HUDs sample selected from ALFALFA without consideration of environment. First we present the properties of the overall sample, showing that the optical and \hi \, properties of our HUDs are similar to those of \citetalias{leisman17a} -- extremely blue, with very narrow velocity widths. We then discuss the environment of the HUDs in comparison to typical ALFALFA galaxies, demonstrating that all metrics we use here suggest that the HUDs and comparison samples are not in significantly different environments.

\subsection{Properties of extended HUDs sample}
\label{sec:sample_properties}

\protect\begin{figure}
    \centering
    \includegraphics[width=\columnwidth]{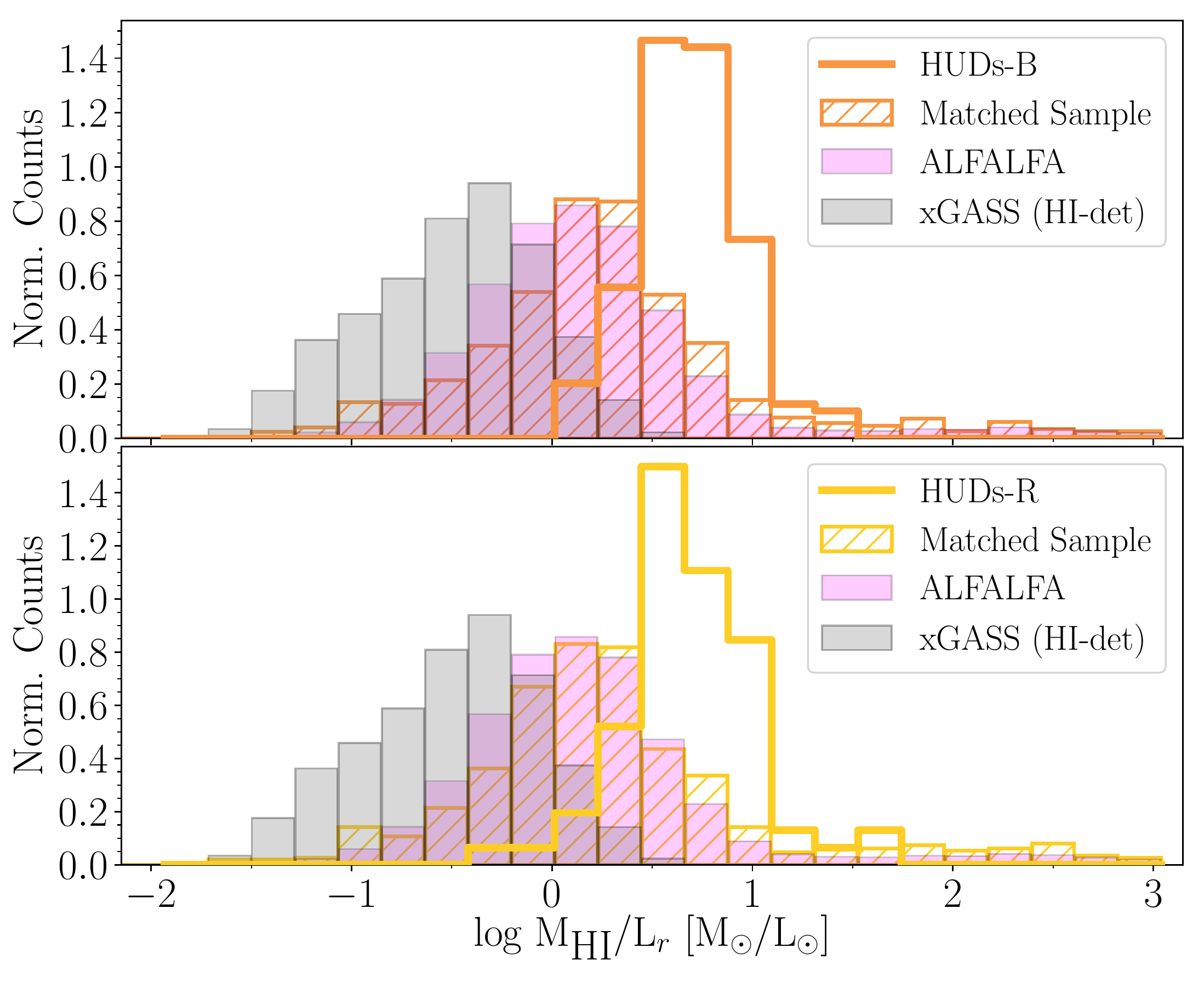}
    \caption{\protect\mhi/{L$_r$} distributions
    are shown for the
    HUDs-B and HUDs-R samples and their matched comparison samples
    (respectively). Also shown are the full ALFALFA
    (\protect\hi-selected) and xGASS (\protect\mst-selected)
    samples.}
    \label{fig:sample_hist}
\protect\end{figure}

As discussed in Section~\ref{sec:sample}, our sample presented here differs from the sample in \citetalias{leisman17a} in two main aspects. First, our sample is based on the ALFALFA 100\% catalog instead of the ALFALFA 70\% catalog, and second, it does not include an environmental isolation criteria, so it includes HUDs from all environments probed by ALFALFA. Thus we compare the effect of these differences  on the sample properties.

\begin{figure}
    \centering
    \includegraphics[width=0.95\columnwidth]{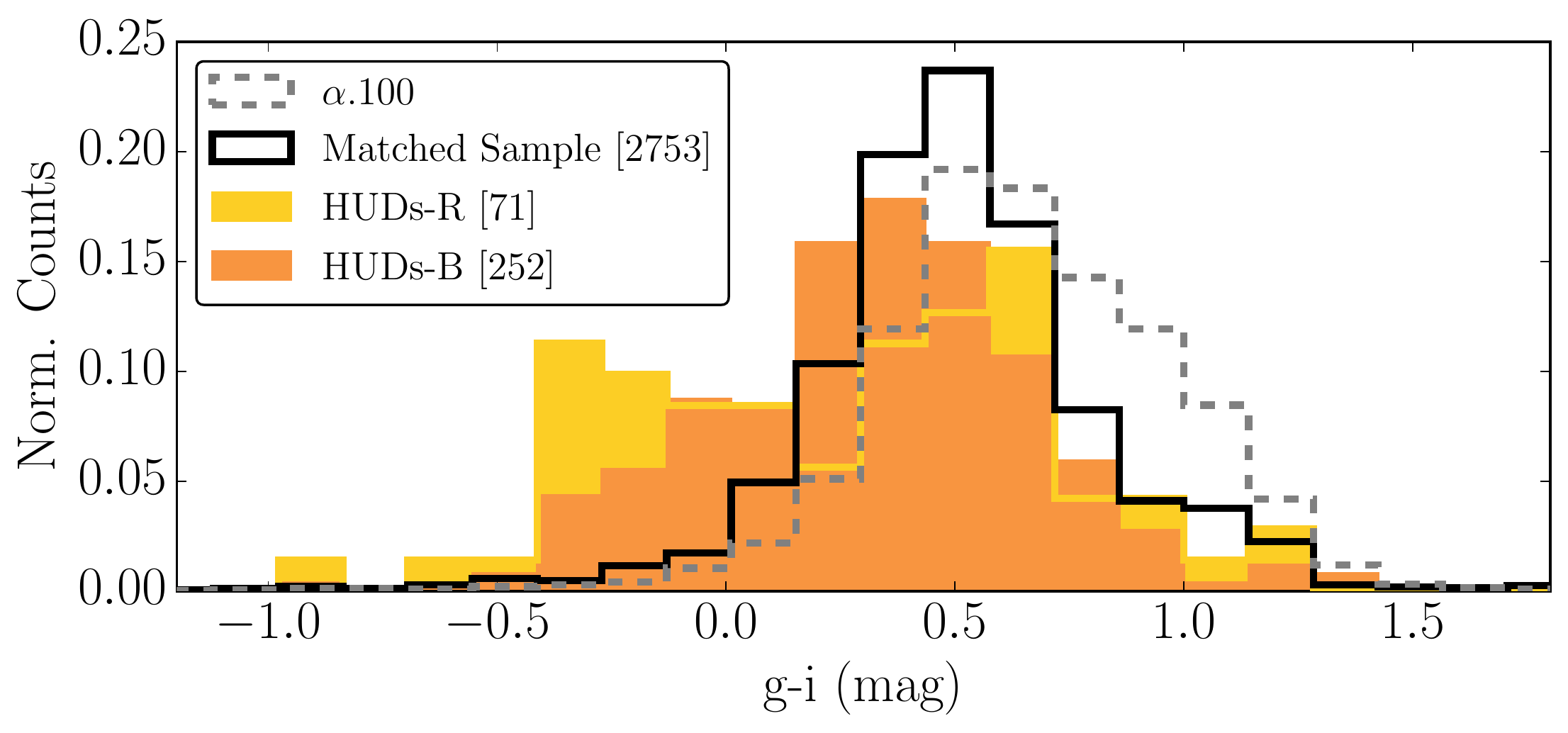}
    \includegraphics[width=0.95\columnwidth]{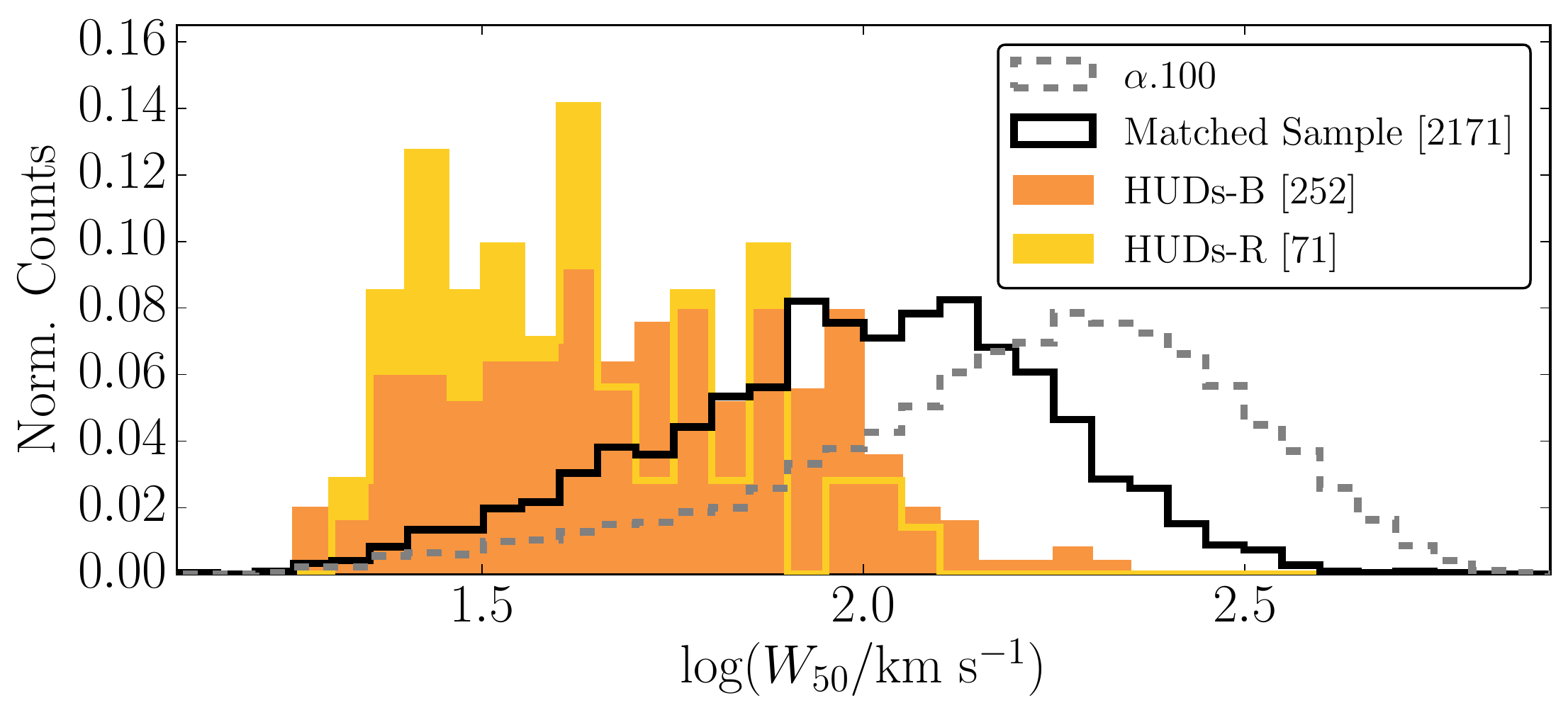}
    \caption{\textit{Top:} The $g$-$i$ colour distribution of HUDs, compared with the matched sample (for HUDs-B) and the 100\% ALFALFA catalog. Both HUDs samples are bluer than the matched sample.
    \textit{Bottom:} \hi\ velocity width distribution of HUDs, compared with the matched sample and the 100\% ALFALFA catalog. ALFALFA HUDs of all environments have narrower velocity widths than other ALFALFA galaxies of similar \hi\ mass.}
    \label{fig:hud_color_width}
\end{figure}

Figure~\ref{fig:sample_hist} includes the distributions of the \mhi \, to $r$ filter luminosity ratio for the HUDs samples and their matched comparison samples. The comparison samples are somewhat less gas rich as they are designed to be matched to the \hi \, properties of the HUDs without any optical constraints. This means the comparison sample galaxies are typically more luminous and less gas rich than the HUDs.   
In these histograms we use L$_r$ instead of stellar mass because the optical photometry of the HUDs has large uncertainties which only become worse when using a colour-dependent mass-to-light ratio. We also include the full ALFALFA sample (of \hi-selected galaxies) as well as the low-mass extension of the GALEX Arecibo SDSS Survey \citep[xGASS,][]{catinella18}, which is a stellar mass-selected sample of galaxies with deep \hi \, observations. These other samples demonstrate the strength of our selection method, as the HUDs are even more gas rich than the (already quite gas rich) ALFALFA sample.

The top panel of Figure~\ref{fig:hud_color_width} shows the distribution of the colour of our measured HUDs in comparison with the full ALFALFA sample, the ALFALFA sample restricted to the same distance limits as the HUDs sample, and the mass- and distance-corrected ALFALFA comparison samples discussed in Section~\ref{sec:ref_sample}. The previous finding that HUDs are bluer than typical ALFALFA galaxies \citepalias{leisman17a} appears to hold for our extended sample.  We find a mean $g$-$i$ colour for the HUDs-B sample of 0.31 mag, with a standard error on the mean of 0.02, compared with 0.45$\pm$0.02 from \citetalias{leisman17a}, and 0.8$\pm$0.1 from \cite{vandokkum15a}. Though our sample includes a significant number of sources in modestly high density environments, the sources still appear to be bluer than the typical ALFALFA galaxy of similar \hi\ mass. We note, however, that while the comparison sample has been matched in \hi\ mass and distance, as noted in Figure \ref{fig:sample_hist}, the HUDs have elevated \hi\ mass to stellar mass ratios, so the stellar masses of comparison galaxies are larger on average.

Further, we emphasize that low signal to noise optical photometry of these poorly detected sources means there is significant uncertainty in any one given colour measurement, and that the result is somewhat sensitive to the SDSS background subtraction. Thus, a full investigation of the colour of HUDs will be better done with deeper optical data.

The bottom panel of Figure~\ref{fig:hud_color_width} shows the distribution of observed \hi \, line width of HUDs compared with the overall ALFALFA sample and the matched comparison sample.  We find good agreement with the result from \citetalias{leisman17a}, that HUDs have significantly narrower velocity widths than sources of similar mass in ALFALFA, even in the larger, more environmentally varied sample.  
A KS test between the higher surface brightness comparison sample and the HUDs sample gives a p-value of  $3.2 \times 10^{-38}$. This result may be indicative of sources that are rotating too slowly for their baryonic mass, and/or of high halo angular momentum as suggested by \citetalias{leisman17a}. However, there may be an inclination-dependent selection effect such that more face on sources will tend to have lower surface brightnesses and be more likely to be selected in our visual examination process. Until deeper optical data are available, this visual inspection remains a necessary part of the selection process. A more significant future exploration is needed to better understand the possible effects of inclination on the observed properties of HUDs.

\begin{figure}
    \centering
    \includegraphics[width=0.95\columnwidth]{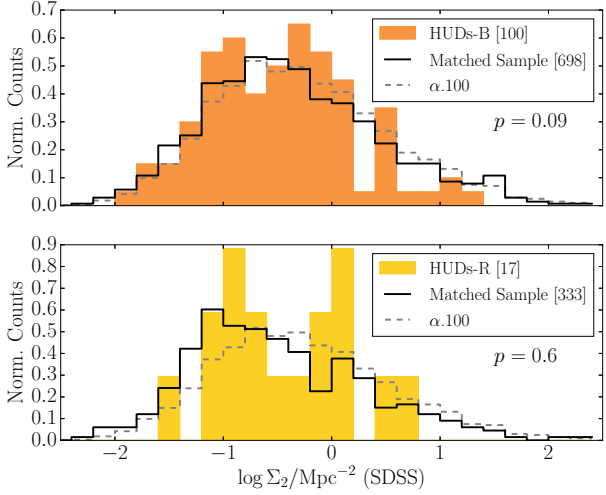}
    \caption{The distribution of 2nd nearest neighbour density in the SDSS spectroscopic reference catalogue for the HUDs-B sample (filled orange bars), the HUDs-R sample (filled yellow bars), the distance- and mass-matched \hi \, sample (black steps), and all ALFALFA objects in the same distance range (dashed grey steps).}
    \label{fig:hud_env_nn_sdss}
\end{figure}

\begin{figure}
    \centering
    \includegraphics[width=0.95\columnwidth]{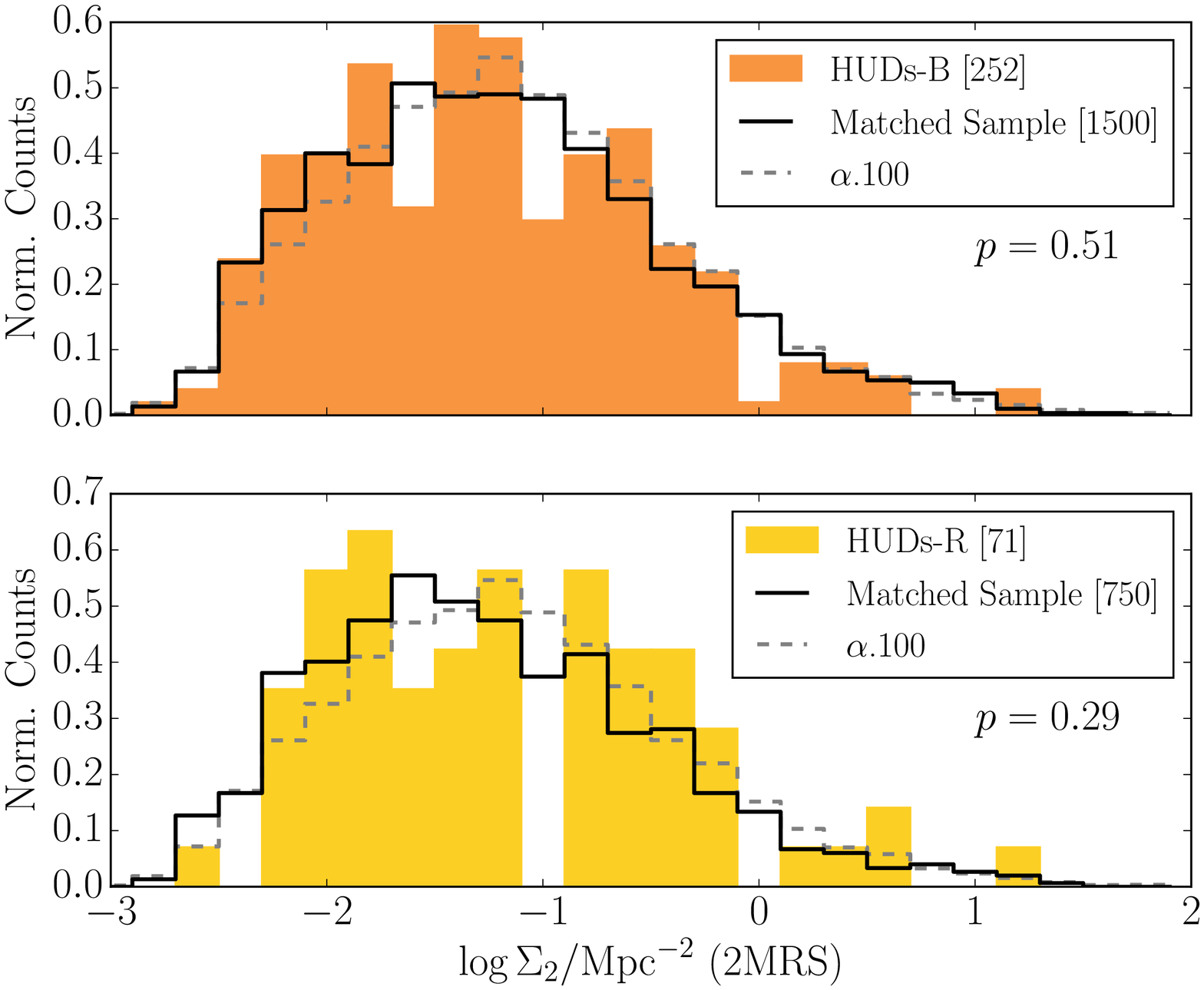}
    \caption{The distribution of 2nd nearest neighbour density in the 2MRS reference catalogue for the HUDs-B sample (filled orange bars), the HUDs-R sample (filled yellow bars), the distance- and mass-matched \hi \, sample (black steps), and all ALFALFA objects in the same distance range (dashed grey steps).}
    \label{fig:hud_env_nn_2mrs}
\end{figure}

\begin{figure}
    \centering
    \includegraphics[width=0.95\columnwidth]{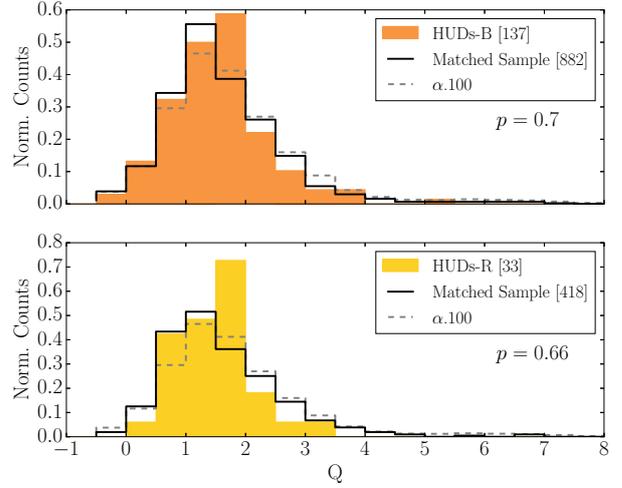}
    \caption{The distribution of the tidal strength parameter ($Q$) calculated from neighbours in the SDSS photometric reference catalogue for the HUDs-B sample (filled orange bars), the HUDs-R sample (filled yellow bars), the distance- and mass-matched \hi \, sample (black steps), and all ALFALFA objects in the same distance range (dashed grey steps).}
    \label{fig:hud_env_q}
\end{figure}

\begin{figure}
    \centering
    \includegraphics[width=0.95\columnwidth]{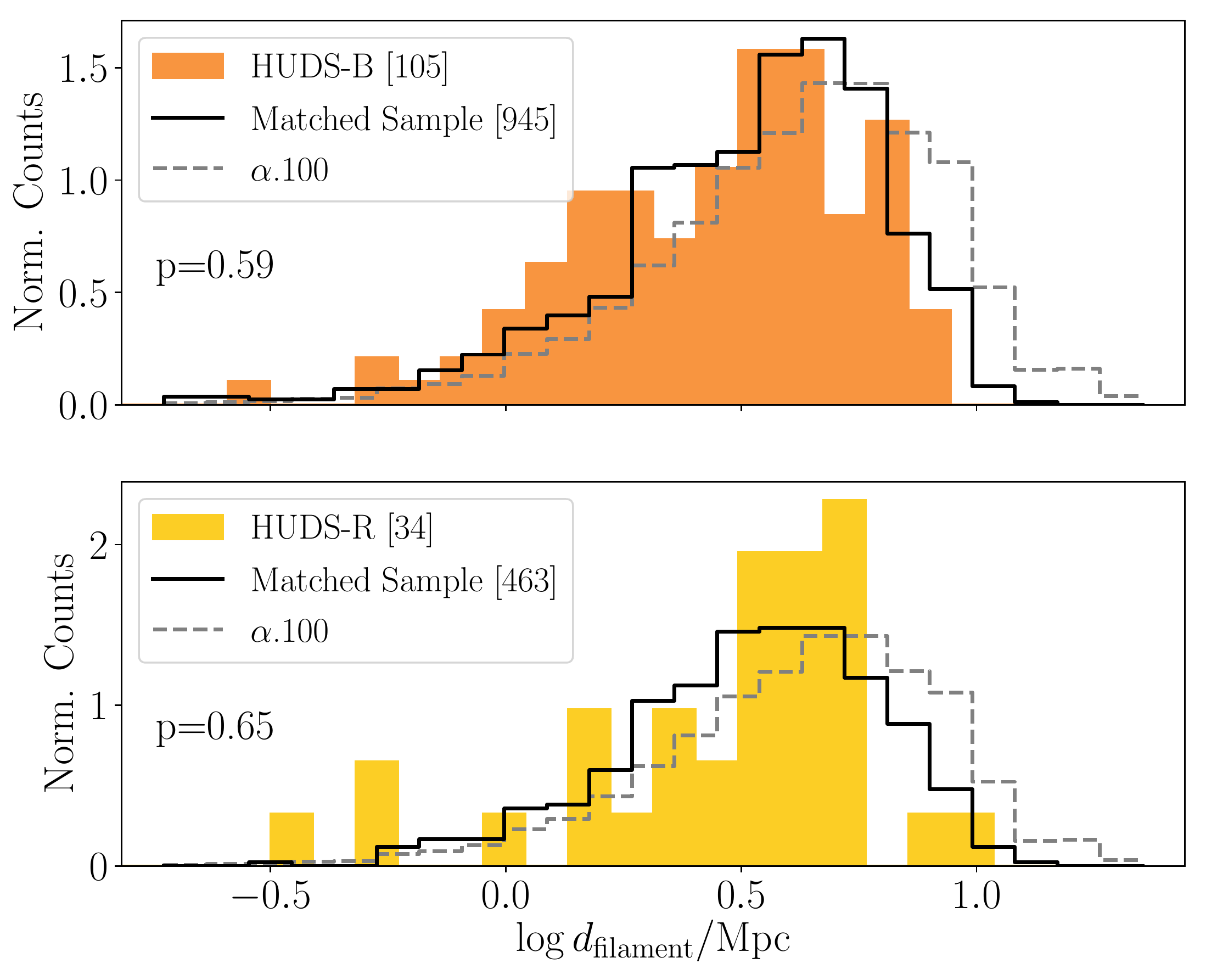}
    \caption{The distribution of $d_\textrm{filament}$  for the HUDs-B sample (filled orange bars), the HUDs-R sample (filled yellow bars), the distance- and mass-matched \hi \, sample (black steps), and all ALFALFA objects in the same distance range (dashed grey steps).}
    \label{fig:hud_env_dfil}
\end{figure}

\subsection{The environment of HUDs}

Figures \ref{fig:hud_env_nn_sdss} \& \ref{fig:hud_env_nn_2mrs} compare the nearest neighbour densities of our two samples of HUDs (HUDs-R and HUDs-B) to their respective matched samples drawn from the general ALFALFA population. In all cases there is no significant evidence of any difference in the distributions of the neighbour densities of HUDs and other similar \hi-selected galaxies that do not meet the surface brightness and size requirements of UDGs. This is true for neighbour densities based on neighbour galaxies selected from both the SDSS spectroscopic sample and the 2MRS sample. For the former, the distribution peaks around $\Sigma_{2} \approx 0.3$, indicating that it is typically probing a galaxy's surrounding environment on a scale of about 1.5~Mpc; whereas for 2MRS the peak is at $\Sigma_{2} \approx 0.03$, which corresponds to a length scale of about 5~Mpc. Thus, these metrics represent measures of the local to intermediate scale environment of the target galaxies, but neither probes the immediate neighbours. This is mostly due to the requirement that the spectroscopic catalogue be volume limited, in order to prevent a distance bias occurring in our measurements of environment, which has the unavoidable side effect of eliminating many fainter galaxies from the reference catalogue.

The tidal $Q$ parameter is a purely photometric measure of the gravitational impact of neighbouring galaxies within a projected distance of 1~Mpc. Therefore, unlike neighbour density it is, by definition, a measurement of a galaxy's immediate environment. If a close neighbour that can exert a strong tidal force was a prerequisite for forming an HUD, then we would expect to see evidence for that in this environment metric. Figure~\ref{fig:hud_env_q} again compares the HUDs sample to the matched ALFALFA sample, but now for their distributions of the $Q$ parameter. Again we find no evidence that the environment of HUDs is different from the matched comparison sample. While this metric suffers from greater scatter than those based on spectroscopic reference catalogues, the number of available sources is considerably larger than for the SDSS neighbour density metric, which helps the KS test to be more discerning.

Figure~\ref{fig:hud_env_dfil} shows the distribution of distances to the nearest filament for the HUDs samples and their matched comparison samples. As with earlier metrics, no significant difference is found between the distributions of these samples. Approximately 1\% of HUDs are found within a filament (i.e., $<0.5$~Mpc from a spine); the same fraction of galaxies from the comparison samples are also found within a filament. On larger scales, 23\% of the HUDs (and 20\% of the comparison samples) are found within 2~Mpc of a filament spine -- this is a moderate difference with a relatively small sample of $\sim$30 HUDs and $\sim$300 galaxies in the comparison sample. Overall, the HUDs samples (both when combined and when treating the HUDs-R and HUDs-B samples separately) do not show any significant preference to be closer or farther from filamentary structures than the matched comparison samples.

\begin{figure}
    \centering
    \includegraphics[width=0.95\columnwidth]{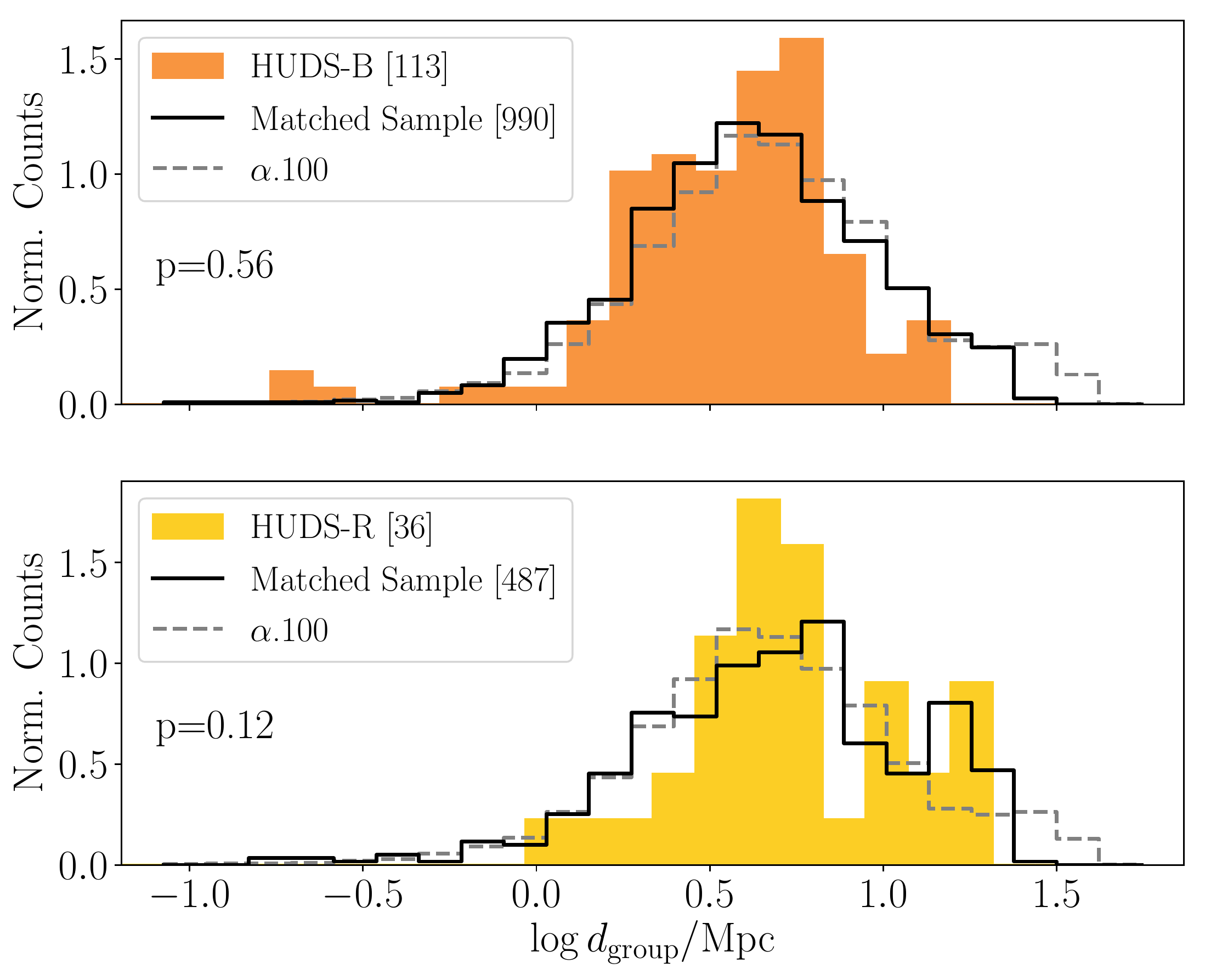}
    \caption{The distribution of $d_\textrm{group}$ for the HUDs-B sample (filled orange bars), the HUDs-R sample (filled yellow bars), the distance and mass matched \hi \, sample (black steps), and all ALFALFA objects in the same distance range (dashed grey steps).}
    \label{fig:hud_env_dgrp}
\end{figure}

Finally, Figure~\ref{fig:hud_env_dgrp} shows the distributions of distances to the nearest group/cluster for our HUDs samples and the comparison samples. The HUDs show no strong differences in distances from groups/clusters than the comparison samples. On average, 52\% of the HUDs (and 57\% of the comparison samples) are found within 5~Mpc of a group/cluster, which is lower than the $\sim$70\% that are found within that same distance of a filament. 
Intriguingly, the HUDs-R and their comparison sample show the most significant statistical difference among all of our environmental metrics -- a p-value of 0.12 in the sense that HUDs may be slightly farther from groups than their comparison sample. While not strongly significant, this may suggest that HUDs-R galaxies prefer to be farther from groups/clusters than galaxies in their comparison sample.

We also considered whether any of our samples (HUDs or matched samples) showed different underlying relationships between environmental metrics and \mhi, which would not be visible in our simple comparison of one-dimensional histograms. These two-dimensional tests confirmed the results shown in our histograms, as the HUDs-B and HUDs-R samples occupy the same regions of parameter space as their matched comparison samples when plotting each environmental indicator against \mhi.  

\section{Discussion}
\label{sec:discussion}

Using a wide variety of environmental metrics, HUDs do not distinguish themselves from the \hi-selected non-ultra diffuse galaxies of our comparison samples. In other words, environment does not seem to play a role in determining which \hi-rich galaxies are ultra diffuse. 
This argues for environmentally independent UDG formation channels that can produce HUDs like these without requiring any interactions beyond those typically experienced by low mass, gas rich galaxies generally found outside of cluster environments. 
The UDGs found in clusters populate the same parameter space as HUDs on the size-luminosity plane as shown in Figure~1 of \citetalias{leisman17a}, although their stellar and gas contents are quite disparate.   
The existence of HUDs with large gas reservoirs 
but unusually small stellar masses   
supports the formation scenario, wherein UDGs are dwarf galaxies with inefficient star formation that live in the high spin tail of the galaxy angular momentum distribution \citep[e.g.,][]{amorisco16a}. HUDs may someday fall into a cluster and be transformed into UDGs similar to those found in clusters, but they are already diffuse even in the absence of strong environmental effects.

While this work has shown that HUDs and their matched comparison samples of \hi-detected galaxies exhibit no differences in environmental preferences, there are other significant differences between these two populations. As seen in Figures~\ref{fig:sample_hist} and \ref{fig:hud_color_width}, the both the HUDs-B and HUDs-R samples are more gas rich, are optically bluer, and have narrower \hi \, line profiles than their matched comparison samples.

In the following subsections, we discuss the effects of our relatively small sample sizes on the robustness of our conclusions, and we consider how their basic  properties (like colour and \hi \, line width) of the HUDs compare with larger \hi-selected samples of galaxies.

\subsection{Required sample size}

We demonstrated in Section \ref{sec:metric_nnden} that even with our modest sample size, a shift in environment on a similar level to that between the SDSS red and blue galaxy populations would be confidently detected. The absence of a clear difference between the HUDs population and the rest of the \hi-selected population (at equivalent \hi \ mass) therefore indicates that there is no strong correlation between a galaxy with extremely low surface brightness and its environment (under the prior condition that it is \hi \, bearing). However, finding that the environment of HUDs is consistent with that of the general \hi \ population is perhaps not surprising given that, by definition, HUDs are \hi \, rich, and properties such as colour correlate much more weakly with environment for \hi-selected galaxies than optically selected galaxies. In other words, this null result may be simply a manifestation of the fact that most \hi-rich galaxies are in relativity low density environments.

\begin{figure*}
    \centering
    \includegraphics[width=0.95\columnwidth]{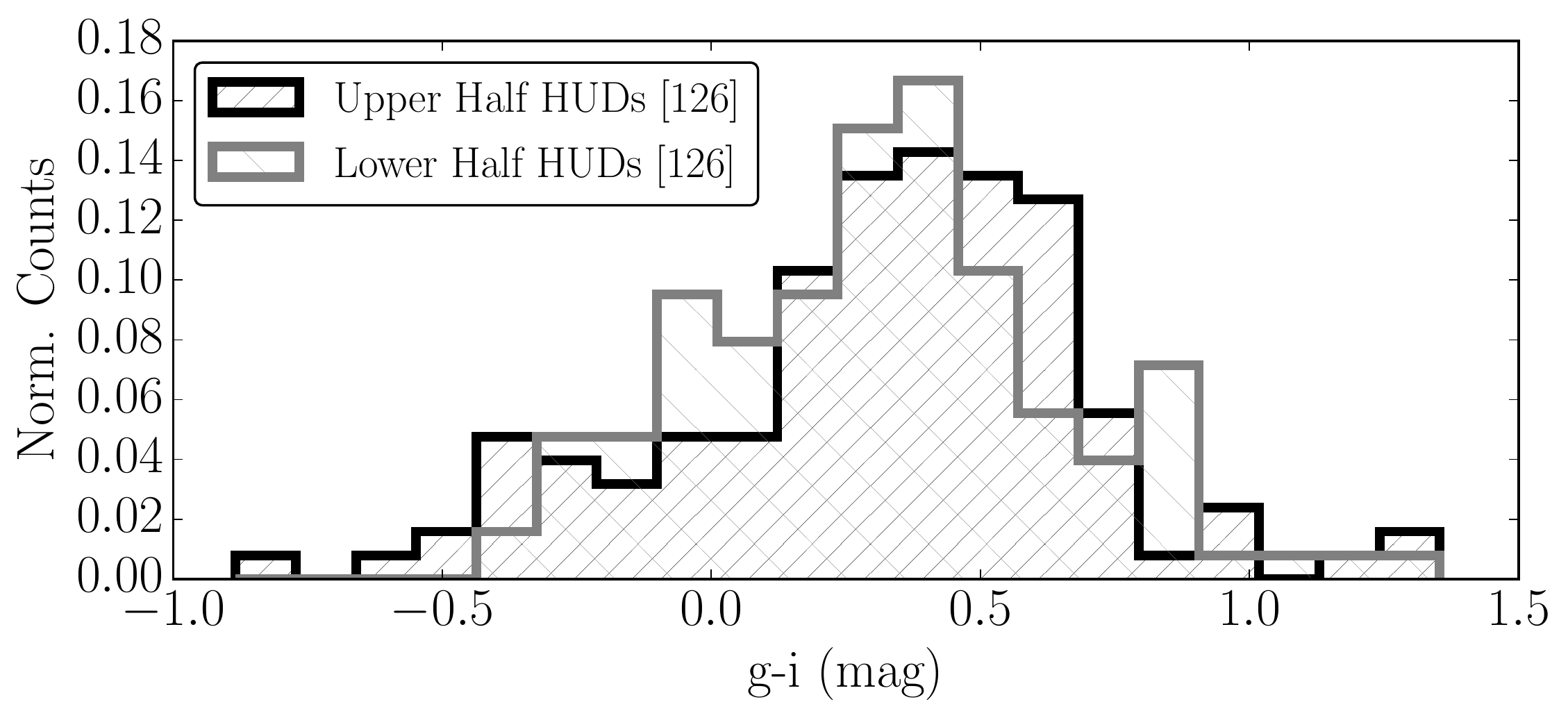}
    \includegraphics[width=0.95\columnwidth]{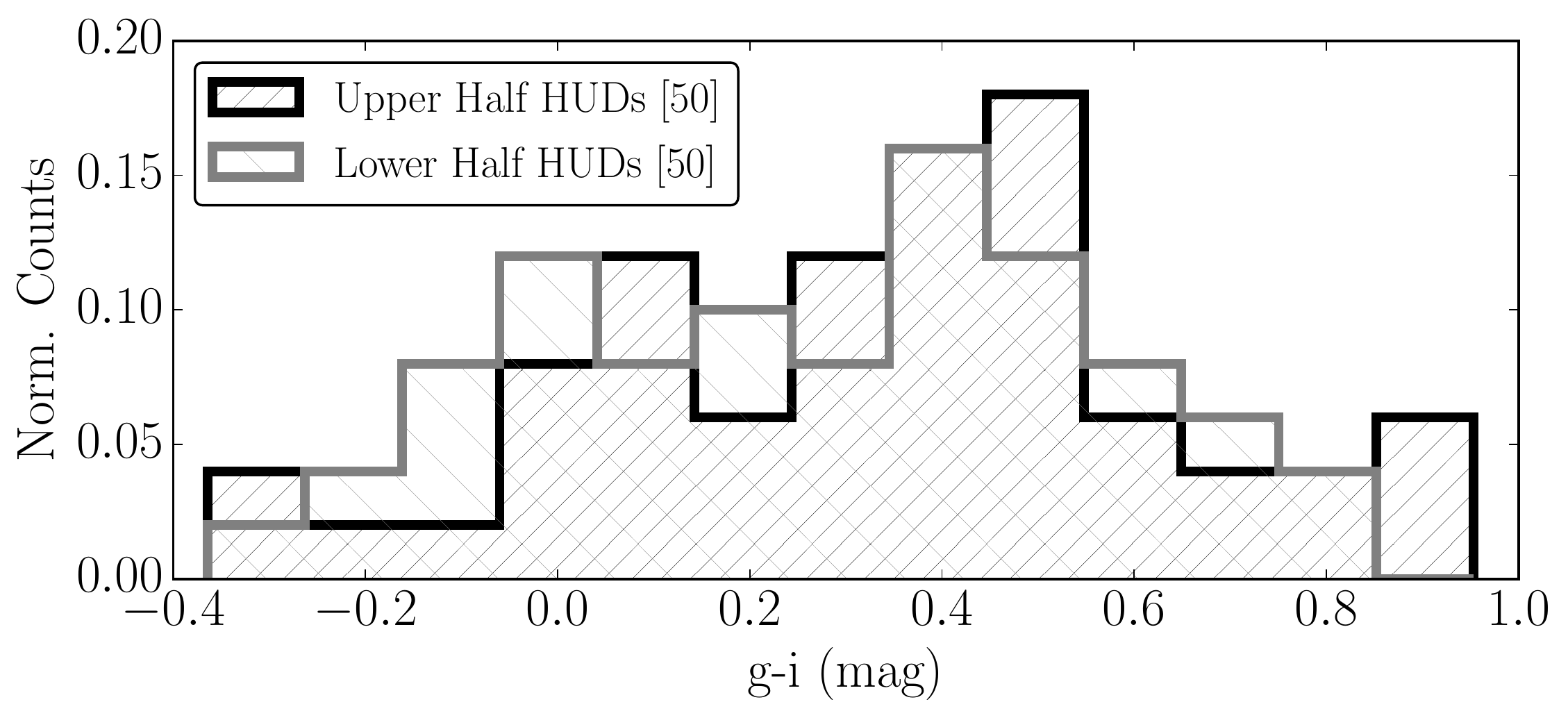}
    \includegraphics[width=0.95\columnwidth]{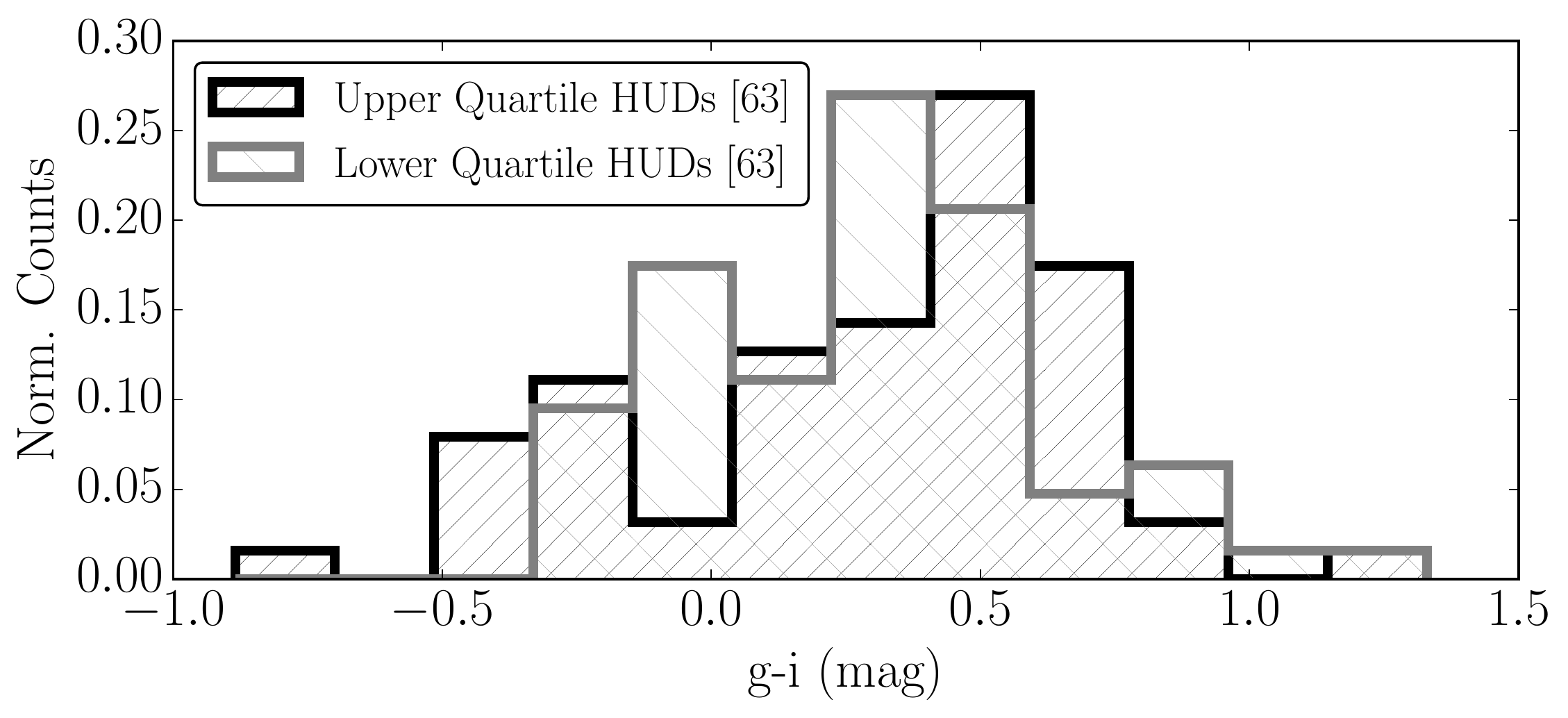}
    \includegraphics[width=0.95\columnwidth]{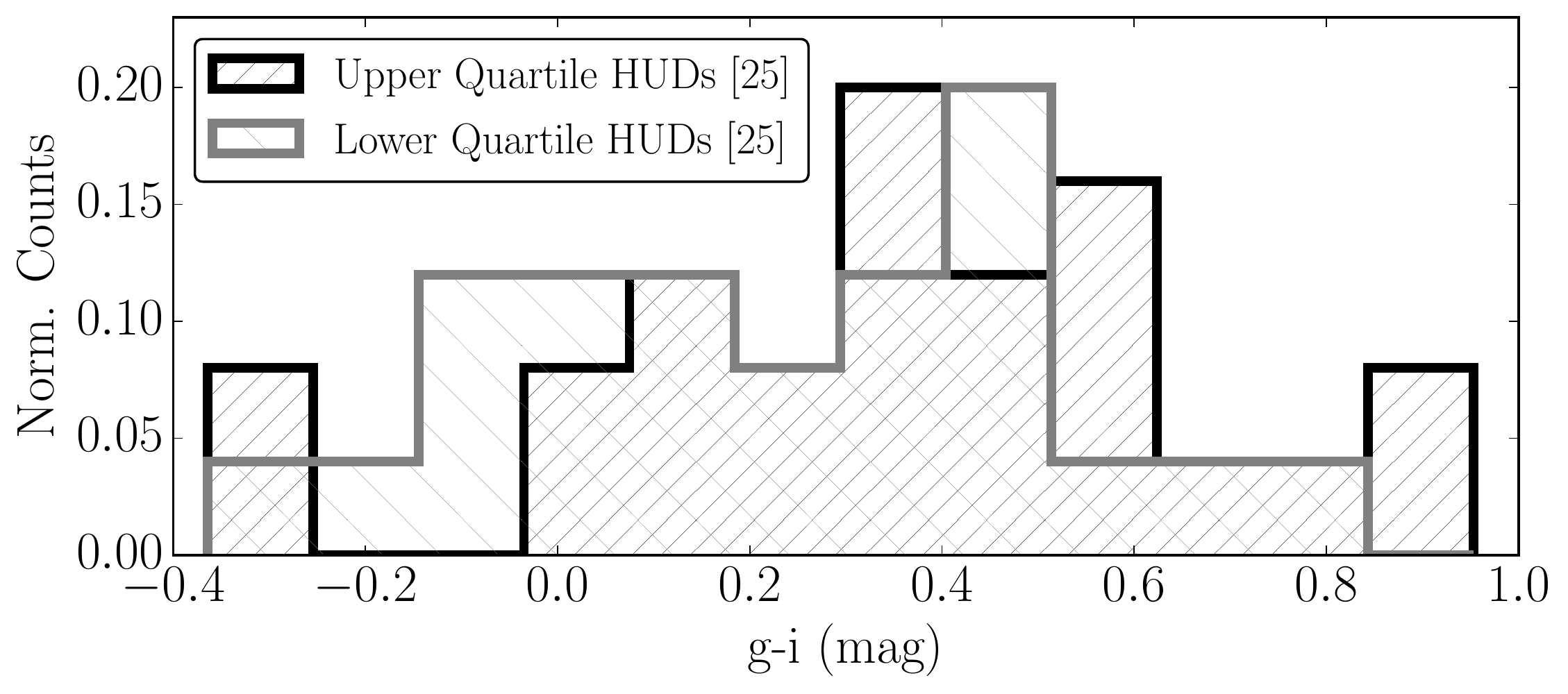}
    \caption{\textit{Left:} Comparison of colour distribution of HUDs in the most and least dense environments, as defined using the 2MRS 2nd nearest neighbour. The top panel shows HUDs in the most dense and lease dense halves, while the bottom panel compares the HUDs in the lowest density quartile to the highest density quartile. The HUDs show no difference in colour as a function of environment (KS test p = 0.70 and 0.18 for halves and quartiles respectively), despite having a statistically large enough sample to potentially see a difference (randomly  drawn samples of 126 sources from the highest and lowest half matched 2MRS samples have a geometric mean p-value of 0.011, and random samples of 63 source drawn from the highest and lowest matched quartiles have a geometric mean p-value of 0.0048).
    \textit{Right:} Same as the left hand panels, except using the SDSS 2nd nearest neighbour definition of environment. No difference is visible between the samples (p-values of 0.68 and 0.41 for half and quartile comparisons), but no significant difference is necessarily expected for the smaller sample sizes (geometric mean p-values of matched samples gives 0.12 and 0.06 for halves and quartiles respectively).}
    \label{fig:HUDcolorCompare}
\end{figure*}

To assess this more quantitatively we took the full $\alpha$.100 population in the distance range $25 <$ D/Mpc $< 120$ that has SDSS photometry and split it into a red ($g-i > 0.85$) and a blue ($g-i < 0.85$) sample, as we did for the SDSS galaxies. The supremum separation between the CDFs (the KS-statistic) of these two populations for our various environmental metrics took values ranging from approximately 0.1 to 0.05, as opposed to approximately 0.3 for the SDSS red and blue samples. To confidently (at 2-$\sigma$ level) detect this difference we estimate that samples with $\mu > 200$ or $\mu > 700$ would be required, respectively.  
For our most sensitive environment metric, changes equivalent to those of the \hi-selected red and blue populations would be detectable at marginal significance in the HUDs-B sample, but not the HUDs-R sample, while for the less sensitive metrics, it would not be possible with either sample.

From this analysis we can conclude that the environment of HUDs is not exceptional in comparison to the general \hi-selected population, and if their properties were to be caused by environmental effects, then those effects would have to be at or below the level that typically exists between \hi-bearing sub-populations.

\subsection{Colour differences between HUDs in different environments}
As discussed above, ALFALFA galaxies follow typical environmental trends with colour, and our environment metrics successfully show these differences. Though HUDs appear to be bluer than typical ALFALFA galaxies of similar \hi \, mass (see Section~\ref{sec:results})  
we still might expect that HUDs in different environments show a trend with colour.   
We explore this possibility using the nearest neighbour 2MRS and SDSS environmental metrics (Section~\ref{sec:metric_nnden}). We divide the HUDs sample into halves and quartiles of environmental density, and then compare the colour between the upper half and the lower half distributions, and between the uppermost quartile and the lowermost quartile distributions. 

Figure \ref{fig:HUDcolorCompare} shows the normalized distributions for each of these comparisons, showing that there appears to be no significant difference in colour between HUDs in high density environments and UDGs in low density environments. Specifically, the left hand panels show the comparison between HUDs in high- and low-density environments as defined by the second nearest neighbour in the 2MRS reference sample. The KS test for the comparison of the HUDs in the two half samples based on this 2MRS metric gives p = 0.704, while the quartile comparison gives p = 0.181. The right hand panels compare the colours of HUDs in low- and high-density environments as defined by the second nearest neighbour in the SDSS reference sample, showing similar results: the KS test for the half comparison results in p = 0.678, and for the quartile comparison p = 0.414. 

However, it is not obvious that these environment metrics are sufficiently sensitive to detect environmental differences between two samples of 25, 50, 63, or 126 galaxies (the number of sources in the two samples in the case of SDSS quartile, SDSS half, 2MRS quartile, and 2MRS half comparisons respectively). To estimate this sensitivity, we took 30,000 random samples from the 
full 2MRS and SDSS half and quartile samples that matched the size of the HUDs  
half and quartile samples.  
When comparing sources in the upper/lower halves and quartiles of the 2MRS samples, significant separations are found between  
the populations with 126 and 63 sources in each sample. The geometric mean p-value of the 30,000 randomly drawn samples of 126 sources from the highest and lowest half matched 2MRS samples is 0.011, while the geometric mean p-value of  the random samples of 63 source drawn from the highest and lowest matched quartiles is 0.0048. 
Using the SDSS 2nd nearest neighbour definition of environment gives less clear results due to the smaller sample sizes (since SDSS covers only a portion of the sky). We find that we expect to see a significant difference between the samples, if one exists, only $\sim$10\% (for quartiles) and $\sim$6\% (for halves) of the time (geometric mean p-values of matched samples gives 0.12 and 0.06 for halves and quartiles respectively).

This result may suggest that HUDs do not show the same trends with environment as typical ALFALFA galaxies, lending support to the case that their low surface brightness is not driven by any environmental process.
However, we caution that the statistical uncertainty in the colour measurements of the HUDs sample is significantly higher than in the full ALFALFA sample -- thus, the lack of trend may simply be an effect of poor data. Future, deeper survey data will be able to better explore this.

\subsection{Number density of HUDs}

Following the methodology of \citet{jones18c} we recalculate the cosmic number density of HUDs using this new sample. Our sample expands on that of \citetalias{leisman17a} by using the 100\% ALFALFA catalog and by relaxing the original environmental isolation criteria of being separated by 350 kpc in projection from another galaxy of similar redshift. We still retain the requirement to be separated from bright stars by at least 10\arcmin \ and to be within the SDSS footprint. We find a consistent result of $(1.2 \pm 0.2) \times 10^{-3}$ HUDs per Mpc$^{3}$, however the increased sample size has reduced the uncertainty by a factor of 3 from the previous work. The fact that this value is consistent with that of \citet{jones18c} is further validation that HUDs reside in environments similar to those of other \hi-selected galaxies of the same mass, as this was a fundamental assumption of the analysis in that paper.

Due to the improved precision of this measurement it is now the case that the uncertainty in selecting HUDs which fully meet the criteria to be classified as UDGs (given their poor photometry) is one of the key limitations, which is not included in the error estimate quoted.

\section{Conclusions}
\label{sec:conclusion}

In this work we have expanded on the efforts of \citet{leisman17a} and selecting two samples of HUDs (restrictive and broad) from SDSS and ALFALFA (100\%) survey observations, now with no environmental restriction included in the selection process. We also selected matched comparison samples for each HUDs sample based on their \hi \, masses and distances, to test for differences in the environmental preferences of HUDs.

Our new sample more than doubles the sample size of known ALFALFA HUDs, both due to an increase in the number of ALFALFA galaxies (we use the full, 100\% ALFALFA catalog), and the removal of environmental restrictions.
We find that the galaxies in our HUDs-R and HUDs-B samples have similar properties to the samples in \citet{leisman17a}, despite having no isolation restrictions. The sample still appears bluer than typical \hi\ selected galaxies of similar \hi\ mass, and have significantly narrower velocity widths, potentially indicative of slower rotational velocities. The sources in the sample are also \hi\ rich, relative to both \hi\ and optically selected samples. That these results hold suggests that environment is not a major factor in these trends.

We additionally find no significant differences between the HUDs and their matched comparison samples using the following environmental metrics:
\begin{itemize}
    \item density of 2nd nearest neighbours in SDSS
    \item density of 2nd nearest neighbours in 2MRS
    \item tidal parameter, $Q$
    \item distance from the nearest filament
    \item distance from the nearest galaxy group/cluster
\end{itemize}

\noindent
In short, environment seems to have no impact on which \hi-rich galaxies are UDGs. 

Our increased sample size results in consistent measurement of the number density of HUDs with \cite{jones18c}, but with reduced uncertainty, again suggesting that HUDs resided in similar environments to other \hi\ selected galaxies.

This environmentally independent behaviour is consistent with a formation scenario wherein UDGs evolve slowly because of low star formation efficiency and do not require an interaction with a cluster to become diffuse. The HUDs in our sample may be the field population of UDGs -- if they fell into a cluster they could have their gas stripped and become similar to other cluster UDGs. However, in their present state they may represent the field component of the UDG population, with cold gas reservoirs and blue optical colours.
Deeper optical imaging and \hi\ synthesis data of this larger sample will be essential in further understanding the formation histories and evolutionary pathways of these enigmatic sources.

\section*{Author statement and acknowledgements}

This work was the result of a highly collaborative effort involving near equal effort from all authors. The idea was originally conceived in discussions between S.J., M.G.J., and L.L. We  describe other contributions to the paper below:
Original Draft:  S.J., M.G.J., L.L.;
Review and Editing: S.J., M.G.J., L.L.;
Sample Selection, Optical Analysis: L.L.;
Comparisons with colour and velocity width: A.W. as part of an extensive senior research project;
2MRS/SDSS Environmental Analysis: M.G.J.;
Group/Filament Environmental Analysis: S.J.;
Funding Acquisition: S.J.
S.J. is lead and corresponding author.

We thank the anonymous referee for their helpful comments which have improved this work. 
The authors acknowledge the work of the entire ALFALFA
collaboration in observing, flagging, and extracting sources. 
We also acknowledge the UWA Research Collaboration Grant, which supported our visits between institutions, and  we gratefully acknowledge fruitful discussions with Martha Haynes, John Salzer, Barbara Catinella, and Luca Cortese.

S.J. acknowledges partial support  from  the  Australian Research Council's Discovery Project funding scheme
(DP150101734).
M.G.J. is supported by a Juan de la Cierva formaci\'{o}n fellowship and also acknowledges support from the grant AYA2015-65973-C3-1-R (MINECO/FEDER, UE).
This work has been supported by the Spanish Science Ministry ``Centro de Excelencia Severo Ochoa'' program under grant SEV-2017-0709.  
The ALFALFA team at Cornell is supported by NSF grants AST-0607007 and AST-1107390 to Riccardo Giovanelli and Martha P. Haynes and by grants from the Brinson Foundation. 

This work is based in part on observations made with the Arecibo Observatory, which is operated by SRI International under a cooperative agreement with the National Science Foundation (AST-1100968), and in alliance with Ana G. M\`endez-Universidad Metropolitana, and the Universities Space Research Association. 

This research has made use of NASA's Astrophysics Data System, and also the NASA/IPAC Extragalactic Database (NED), which is operated by the Jet Propulsion Laboratory, California Institute of  Technology, under contract  with  the  National Aeronautics  and Space Administration. This research has also made extensive use of the
invaluable Tool for OPerations on Catalogues And Tables
\citep[TOPCAT][]{taylor05a}.

This research used data from the Sloan Digital Sky Survey, funded by
the Alfred P. Sloan Foundation, the participating institutions, the
National Science Foundation, the U.S. Department of Energy, the
National Aeronautics and Space Administration, the Japanese
Monbukagakusho, the Max Planck Society, and the Higher Education
Funding Council for England.




\bibliographystyle{mnras}
\bibliography{mybib} 


\clearpage
\onecolumn

\appendix
\section{Optical properties of HUDs samples}
\label{sec:optical}

The following table (available online-only) gives the key observed quantities for the HUDs-R and HUDs-B samples, as well as their best-fitting surface brightness profile values in the SDSS $g$ filter. These profile fits are somewhat uncertain (as reflected in the uncertainties on $\mu_0$ and r$_e$ in parentheses), but all HUDs in our samples have been visually confirmed as low surface brightness sources.

\vspace{1cm}

\tablehead{
 \multicolumn{9}{l}{Table~A: Observations and derived optical properties of HUDs-R and HUDs-B samples} \\
   Sample & AGC  & RA      & Dec     & D     & M$_\textrm{\hi}$ &
        M$_g$ & $\mu_0$($g$) & r$_e$($g$)  \\
        & & [h:m:s] & [d:m:s] & [Mpc] & [\msun] &
       [mag] & [\magsec] & 
       [$"$]  \\
\hline
}
\begin{supertabular}{llllllllr}
HUDs-R & 103085 & 00:03:57.4 & +05:24:04 & 42.3  &  8.25 & -14.3 & 24.3 (0.1) &  4.9  (0.9) \\
HUDs-R & 102375 & 00:14:59.7 & +02:34:48 & 35.8  &  8.23 & -14.5 & 24.1 (0.1) &  5.8  (0.8) \\
HUDs-R & 103435 & 00:15:21.2 & +01:04:30 & 28.5  &  8.28 & -14.7 & 24.2 (0.1) &  8.3  (1.4) \\
HUDs-B & 102907 & 00:17:46.3 & +28:44:58 & 89.3  &  9.30 & -16.6 & 23.5 (0.1) &  4.6  (0.5) \\
HUDs-B & 102761 & 00:19:43.9 & +30:03:26 & 94.3  &  9.30 & -15.4 & 23.8 (0.1) &  2.9  (0.4) \\
HUDs-R & 103796 & 00:20:39.6 & +06:57:57 & 79.6  &  8.85 & -15.8 & 24.2 (0.1) &  4.9  (0.9) \\
HUDs-B & 103797 & 00:24:07.4 & +05:56:30 & 98.4  &  9.26 & -16.2 & 23.8 (0.1) &  4.1  (0.6) \\
HUDs-R & 748789 & 00:24:32.1 & +15:35:13 & 74.9  &  8.82 & -15.7 & 24.8 (0.1) &  6.7  (1.7) \\
HUDs-B & 105118 & 00:26:26.5 & +13:44:54 & 72.8  &  8.90 & -15.2 & 23.5 (0.1) &  2.9  (0.5) \\
HUDs-B & 102919 & 00:27:56.8 & +29:36:05 & 75.0  &  9.33 & -17.1 & 24.0 (0.1) &  8.7  (1.1) \\
HUDs-R & 100288 & 00:31:45.4 & +02:42:53 & 32.9  &  8.50 & -15.6 & 24.3 (0.1) & 11.4  (1.9) \\
HUDs-R & 105466 & 00:31:54.5 & +02:49:43 & 77.0  &  8.86 & -14.8 & 24.7 (0.2) &  4.1  (1.2) \\
HUDs-B & 104543 & 00:34:04.2 & +36:10:28 & 67.4  &  8.95 & -16.1 & 24.0 (0.1) &  6.4  (0.8) \\
HUDs-B & 102791 & 00:40:57.7 & +31:54:23 & 75.4  &  9.28 & -16.2 & 23.8 (0.1) &  5.4  (0.6) \\
HUDs-B & 104550 & 00:42:25.9 & +35:47:12 & 91.7  &  9.01 & -15.4 & 23.6 (0.1) &  2.8  (0.4) \\
HUDs-B & 102269 & 00:44:13.4 & +26:37:30 & 74.8  &  9.02 & -16.5 & 23.9 (0.1) &  6.5  (0.8) \\
HUDs-B & 104153 & 00:50:36.2 & +32:21:05 & 65.8  &  8.62 & -16.4 & 24.2 (0.1) &  7.9  (1.2) \\
HUDs-B & 105418 & 00:53:17.3 & +04:31:44 & 59.3  &  8.98 & -16.2 & 23.9 (0.1) &  6.8  (0.8) \\
HUDs-B & 104492 & 00:54:07.3 & +32:10:21 & 104.8 &  9.23 & -15.7 & 23.9 (0.1) &  3.2  (0.5) \\
HUDs-R & 105346 & 00:56:17.8 & +11:20:03 & 59.4  &  8.62 & -14.2 & 24.3 (0.2) &  3.4  (1.0) \\
HUDs-R & 101240 & 00:58:39.4 & +31:21:43 & 74.4  &  8.86 & -16.0 & 24.7 (0.1) &  7.4  (1.8) \\
HUDs-B & 103857 & 00:59:13.6 & +06:16:24 & 63.8  &  8.93 & -16.1 & 23.7 (0.1) &  5.7  (0.7) \\
HUDs-B & 748801 & 01:00:52.7 & +16:04:57 & 60.5  &  9.09 & -16.5 & 23.5 (0.1) &  6.8  (0.7) \\
HUDs-B & 116421 & 01:02:32.7 & +05:32:37 & 76.0  &  8.82 & -15.4 & 23.5 (0.1) &  3.2  (0.4) \\
HUDs-R & 113790 & 01:13:02.1 & +27:38:13 & 68.7  &  8.57 & -15.5 & 24.3 (0.1) &  5.3  (0.9) \\
HUDs-B & 114882 & 01:13:25.5 & +06:20:38 & 107.0 &  9.44 & -15.6 & 24.1 (0.1) &  3.3  (0.7) \\
HUDs-B & 115590 & 01:18:40.0 & +35:12:55 & 70.8  &  9.10 & -15.6 & 23.4 (0.1) &  3.5  (0.5) \\
HUDs-B & 113549 & 01:20:23.4 & +00:01:49 & 50.0  &  8.82 & -15.4 & 23.9 (0.1) &  5.8  (0.8) \\
HUDs-R & 114754 & 01:22:29.9 & +08:45:42 & 33.8  &  8.19 & -15.5 & 24.6 (0.1) & 12.4  (3.2) \\
HUDs-B & 110319 & 01:25:16.6 & +14:08:55 & 69.9  &  9.22 & -16.4 & 23.7 (0.1) &  6.0  (0.6) \\
HUDs-R & 114905 & 01:25:18.5 & +07:21:37 & 75.7  &  9.11 & -16.3 & 24.9 (0.1) &  9.0  (2.4) \\
HUDs-B & 115203 & 01:26:22.0 & +32:38:07 & 56.7  &  9.07 & -15.7 & 23.5 (0.1) &  5.0  (0.5) \\
HUDs-B & 748816 & 01:27:56.8 & +14:30:07 & 87.2  &  9.40 & -16.8 & 24.4 (0.1) &  8.0  (1.6) \\
HUDs-B & 114959 & 01:31:18.8 & +11:42:52 & 34.6  &  8.51 & -14.9 & 24.1 (0.1) &  7.2  (1.1) \\
HUDs-B & 115674 & 01:37:16.6 & +32:34:16 & 86.1  &  8.95 & -16.2 & 24.0 (0.1) &  5.2  (0.9) \\
HUDs-B & 115906 & 01:41:40.1 & +16:52:19 & 119.3 &  9.26 & -17.1 & 23.8 (0.1) &  5.0  (0.6) \\
HUDs-R & 114943 & 01:47:06.6 & +07:19:52 & 116.1 &  9.10 & -16.4 & 24.5 (0.1) &  5.1  (1.1) \\
HUDs-R & 113949 & 01:49:38.6 & +30:40:51 & 101.6 &  9.03 & -15.7 & 24.2 (0.1) &  3.9  (0.9) \\
HUDs-B & 116310 & 01:50:00.1 & +12:36:06 & 72.7  &  8.85 & -16.0 & 24.1 (0.1) &  5.8  (0.9) \\
HUDs-R & 111947 & 01:50:00.9 & +28:54:54 & 52.2  &  8.86 & -16.4 & 24.3 (0.1) & 10.5  (1.8) \\
HUDs-B & 115292 & 01:51:39.5 & +17:17:18 & 34.5  &  8.32 & -15.1 & 24.1 (0.1) &  8.0  (1.0) \\
HUDs-B & 116505 & 01:57:20.7 & +03:27:55 & 63.6  &  8.84 & -14.5 & 23.9 (0.1) &  3.1  (0.5) \\
HUDs-B & 114607 & 01:58:07.1 & +00:52:41 & 80.2  &  8.91 & -15.9 & 23.8 (0.1) &  4.4  (0.6) \\
HUDs-R & 114124 & 01:58:45.6 & +24:30:33 & 66.0  &  8.62 & -15.6 & 24.6 (0.1) &  6.5  (1.4) \\
HUDs-B & 110718 & 01:59:42.8 & +10:24:56 & 86.1  &  9.24 & -16.7 & 23.8 (0.1) &  5.9  (0.8) \\
HUDs-B & 124633 & 02:00:26.4 & +21:24:14 & 41.8  &  8.33 & -14.6 & 24.1 (0.1) &  5.4  (0.9) \\
HUDs-B & 122791 & 02:03:28.8 & +26:41:50 & 68.8  &  8.80 & -16.0 & 24.2 (0.1) &  6.3  (1.0) \\
HUDs-B & 125043 & 02:05:02.9 & +21:10:15 & 68.2  &  8.95 & -15.3 & 23.8 (0.1) &  3.9  (0.6) \\
HUDs-R & 125053 & 02:08:42.7 & +19:58:49 & 70.9  &  8.96 & -15.3 & 24.7 (0.2) &  5.7  (1.7) \\
HUDs-R & 122966 & 02:09:29.0 & +31:51:10 & 89.6  &  9.00 & -16.2 & 25.3 (0.2) &  8.8  (3.5) \\
HUDs-B & 123938 & 02:10:46.7 & +01:46:36 & 48.1  &  8.71 & -15.8 & 23.8 (0.1) &  7.1  (0.7) \\
HUDs-B & 125556 & 02:13:08.3 & +12:36:35 & 87.3  &  9.00 & -16.9 & 24.5 (0.1  &  8.7  (1.7) \\
HUDs-B & 122986 & 02:18:04.4 & +32:02:01 & 85.3  &  8.92 & -16.5 & 23.5 (0.1) &  4.6  (0.5) \\
HUDs-R & 125273 & 02:20:23.1 & +20:44:10 & 56.7  &  8.64 & -14.1 & 25.0 (0.2) &  4.7  (1.8) \\
HUDs-B & 124634 & 02:22:05.5 & +18:24:33 & 30.6  &  8.48 & -14.6 & 24.7 (0.1) &  9.3  (2.4) \\
HUDs-B & 123932 & 02:41:08.7 & +01:05:57 & 100.4 &  9.55 & -15.7 & 23.4 (0.1) &  2.5  (0.3) \\
HUDs-B & 124965 & 02:45:58.8 & +33:02:16 & 65.4  &  9.01 & -16.3 & 24.1 (0.1) &  7.5  (1.1) \\
HUDs-B & 125675 & 02:54:17.6 & +05:09:31 & 103.6 &  9.39 & -17.6 & 25.6 (0.2) & 16.1  (8.6) \\
HUDs-B & 123953 & 02:56:03.5 & +01:58:11 & 73.3  &  8.96 & -15.4 & 23.6 (0.1) &  3.4  (0.4) \\
HUDs-B & 125682 & 02:58:22.5 & +04:44:57 & 102.6 &  9.03 & -15.1 & 24.1 (0.2) &  2.6  (0.6) \\
HUDs-R & 132240 & 03:02:54.5 & +36:07:05 & 105.9 &  9.22 & -15.5 & 25.0 (0.2) &  4.8  (1.7) \\
HUDs-B & 749251 & 07:35:00.3 & +26:39:32 & 106.2 &  9.04 & -15.9 & 24.1 (0.1) &  3.7  (0.6) \\
HUDs-B & 174593 & 07:42:31.8 & +10:19:17 & 71.0  &  8.84 & -16.4 & 23.7 (0.1) &  5.8  (0.6) \\
HUDs-R & 749376 & 07:58:08.9 & +25:49:50 & 68.8  &  8.66 & -15.0 & 24.5 (0.1) &  4.6  (1.1) \\
HUDs-B & 188739 & 08:00:30.1 & +12:03:49 & 71.9  &  9.18 & -16.1 & 23.8 (0.1) &  5.3  (0.6) \\
HUDs-R & 181474 & 08:06:12.4 & +15:30:15 & 30.4  &  8.45 & -15.0 & 24.5 (0.1) & 10.2  (1.8) \\
HUDs-R & 749387 & 08:26:56.4 & +24:43:44 & 34.4  &  8.11 & -15.7 & 24.3 (0.1) & 11.2  (1.8) \\
HUDs-B & 188768 & 08:28:37.8 & +14:58:23 & 87.4  &  8.81 & -15.9 & 24.0 (0.1) &  4.4  (0.7) \\
HUDs-B & 749388 & 08:33:11.3 & +24:34:19 & 78.7  &  8.82 & -16.2 & 23.6 (0.1) &  4.7  (0.5) \\
HUDs-B & 189323 & 08:33:37.8 & +31:18:19 & 79.4  &  9.09 & -16.5 & 24.0 (0.1) &  6.5  (0.9) \\
HUDs-B & 749279 & 08:35:08.2 & +26:52:55 & 107.6 &  9.13 & -15.4 & 23.4 (0.1) &  2.1  (0.3) \\
HUDs-B & 749282 & 08:40:22.7 & +27:13:17 & 79.4  &  9.15 & -15.8 & 23.2 (0.1) &  3.1  (0.3) \\
HUDs-R & 189327 & 08:42:11.8 & +30:19:52 & 82.3  &  9.17 & -16.2 & 24.5 (0.1) &  6.8  (1.4) \\
HUDs-B & 189096 & 08:47:03.6 & +00:10:00 & 60.0  &  8.90 & -16.0 & 23.4 (0.1) &  5.1  (0.5) \\
HUDs-B & 189298 & 08:52:26.6 & +30:43:37 & 31.8  &  8.19 & -14.8 & 23.9 (0.1) &  6.9  (0.8) \\
HUDs-B & 189167 & 08:54:10.9 & +28:58:04 & 57.9  &  8.50 & -15.3 & 23.8 (0.1) &  4.6  (0.6) \\
HUDs-B & 182467 & 08:57:05.6 & +09:07:06 & 59.6  &  8.66 & -15.7 & 23.8 (0.1) &  5.4  (0.6) \\
HUDs-B & 198686 & 09:03:48.3 & +31:47:01 & 30.0  &  8.25 & -15.2 & 23.9 (0.1) &  9.0  (1.1) \\
HUDs-B & 198560 & 09:07:20.6 & +29:10:34 & 102.4 &  9.27 & -17.0 & 24.3 (0.1) &  7.2  (1.1) \\
HUDs-B & 198563 & 09:14:40.2 & +28:30:39 & 95.3  &  9.18 & -15.6 & 24.1 (0.1) &  3.7  (0.7) \\
HUDs-B & 198564 & 09:15:55.8 & +29:55:27 & 108.2 &  9.38 & -16.7 & 23.8 (0.1) &  4.6  (0.6) \\
HUDs-R & 749290 & 09:16:01.1 & +26:38:59 & 96.8  &  8.95 & -16.0 & 24.2 (0.1) &  4.5  (0.8) \\
HUDs-R & 198452 & 09:16:41.5 & +06:26:07 & 82.6  &  8.95 & -16.2 & 24.3 (0.1) &  5.9  (1.2) \\
HUDs-B & 193800 & 09:17:44.9 & +10:30:16 & 76.6  &  9.00 & -15.5 & 23.7 (0.1) &  3.6  (0.5) \\
HUDs-B & 198814 & 09:19:47.2 & +33:36:00 & 65.0  &  8.85 & -15.4 & 24.0 (0.1) &  4.7  (0.7) \\
HUDs-B & 198539 & 09:22:25.4 & +03:15:31 & 52.8  &  8.90 & -16.6 & 23.8 (0.1) &  9.0  (0.9) \\
HUDs-R & 749401 & 09:24:20.7 & +25:39:40 & 41.4  &  8.84 & -16.0 & 24.5 (0.1) & 11.6  (2.0) \\
HUDs-B & 198540 & 09:25:13.9 & +03:18:47 & 63.3  &  9.12 & -16.1 & 23.5 (0.1) &  5.3  (0.5) \\
HUDs-B & 198477 & 09:25:20.4 & +03:06:54 & 77.9  &  9.49 & -17.2 & 23.9 (0.1) &  8.4  (1.0) \\
HUDs-R & 198543 & 09:35:21.7 & +02:59:10 & 82.9  &  9.19 & -16.5 & 24.9 (0.2) &  8.9  (2.5) \\
HUDs-B & 193833 & 09:40:13.2 & +09:55:31 & 51.3  &  8.66 & -15.6 & 23.5 (0.1) &  5.1  (0.5) \\
HUDs-R & 191708 & 09:40:27.0 & +00:02:33 & 29.3  &  8.34 & -15.0 & 24.3 (0.1) &  9.8  (1.5) \\
HUDs-R & 198596 & 09:48:07.2 & +16:15:38 & 56.9  &  8.67 & -16.4 & 24.8 (0.1) & 12.3  (3.3) \\
HUDs-B & 198570 & 09:48:15.8 & +28:12:38 & 98.8  &  9.13 & -16.2 & 23.7 (0.1) &  3.8  (0.5) \\
HUDs-B & 198802 & 09:50:46.7 & +30:47:48 & 64.8  &  8.85 & -15.4 & 23.7 (0.1) &  4.1  (0.5) \\
HUDs-B & 198598 & 09:51:03.4 & +16:54:11 & 56.6  &  8.46 & -15.3 & 24.1 (0.1) &  5.4  (0.9) \\
HUDs-B & 191818 & 09:53:02.2 & +07:47:23 & 39.9  &  8.13 & -15.2 & 24.0 (0.1) &  6.7  (0.8) \\
HUDs-B & 193841 & 09:56:27.0 & +10:54:20 & 79.8  &  8.94 & -15.0 & 23.9 (0.1) &  3.0  (0.5) \\
HUDs-R & 193798 & 09:59:27.8 & +16:00:28 & 62.2  &  8.95 & -14.9 & 24.3 (0.1) &  4.5  (0.9) \\
HUDs-B & 208693 & 10:13:28.8 & +18:36:45 & 54.7  &  8.87 & -15.3 & 23.5 (0.1) &  4.1  (0.4) \\
HUDs-R & 201993 & 10:15:59.5 & +06:48:15 & 25.8  &  8.41 & -15.1 & 24.6 (0.1) & 13.7  (3.2) \\
HUDs-B & 208764 & 10:27:16.6 & +20:12:55 & 107.4 &  9.02 & -16.1 & 23.2 (0.1) &  2.7  (0.3) \\
HUDs-B & 208759 & 10:29:07.7 & +20:14:48 & 58.6  &  8.49 & -15.1 & 23.6 (0.1) &  3.8  (0.5) \\
HUDs-B & 208892 & 10:30:26.4 & +32:13:16 & 108.9 &  9.33 & -17.1 & 23.9 (0.1) &  5.6  (0.7) \\
HUDs-B & 202015 & 10:31:54.0 & +12:55:37 & 43.1  &  8.28 & -15.1 & 23.7 (0.1) &  5.4  (0.6) \\
HUDs-R & 205062 & 10:34:02.2 & +15:22:15 & 99.2  &  9.26 & -16.4 & 24.3 (0.1) &  5.5  (1.1) \\
HUDs-B & 749409 & 10:34:25.9 & +24:04:29 & 77.5  &  9.17 & -15.6 & 23.5 (0.1) &  3.5  (0.4) \\
HUDs-B & 208308 & 10:36:19.8 & +06:25:08 & 76.3  &  8.76 & -16.8 & 24.7 (0.1) & 10.5  (2.4) \\
HUDs-B & 749318 & 10:37:52.8 & +27:57:57 & 116.1 &  9.29 & -16.2 & 24.0 (0.1) &  3.8  (0.6) \\
HUDs-R & 208769 & 10:49:53.5 & +21:38:18 & 89.3  &  9.03 & -16.1 & 24.6 (0.1) &  6.2  (1.5) \\
HUDs-B & 749417 & 10:51:31.7 & +25:16:27 & 89.8  &  9.11 & -16.9 & 24.5 (0.1) &  8.3  (1.7) \\
HUDs-B & 205061 & 10:57:27.4 & +09:10:28 & 41.5  &  8.25 & -14.5 & 24.0 (0.1) &  4.8  (0.7) \\
HUDs-B & 208368 & 10:59:25.5 & +08:36:29 & 94.0  &  9.10 & -15.7 & 23.9 (0.1) &  3.6  (0.6) \\
HUDs-B & 215259 & 11:04:32.6 & +16:06:41 & 92.4  &  9.27 & -16.1 & 23.8 (0.1) &  4.2  (0.6) \\
HUDs-B & 749425 & 11:06:39.2 & +24:02:01 & 85.8  &  9.05 & -15.2 & 23.5 (0.1) &  2.6  (0.4) \\
HUDs-B & 215277 & 11:12:07.4 & +12:38:00 & 49.2  &  8.60 & -14.7 & 24.1 (0.1) &  4.5  (0.8) \\
HUDs-B & 215283 & 11:15:18.4 & +12:53:55 & 90.9  &  9.20 & -16.3 & 23.6 (0.1) &  4.3  (0.5) \\
HUDs-R & 219199 & 11:16:35.3 & +04:24:51 & 39.3  &  8.74 & -15.0 & 24.3 (0.1) &  7.3  (1.3) \\
HUDs-R & 219200 & 11:22:20.0 & +03:53:56 & 25.7  &  8.22 & -13.9 & 25.0 (0.2) &  9.6  (3.4) \\
HUDs-B & 219672 & 11:31:31.7 & +21:32:28 & 97.2  &  9.19 & -15.8 & 23.4 (0.1) &  2.8  (0.3) \\
HUDs-R & 219626 & 11:33:48.6 & +18:57:34 & 55.5  &  8.54 & -16.1 & 24.2 (0.1) &  8.3  (1.4) \\
HUDs-B & 219677 & 11:35:55.5 & +20:28:22 & 87.9  &  8.96 & -15.7 & 23.8 (0.1) &  3.5  (0.5) \\
HUDs-B & 219487 & 11:37:44.8 & +28:11:45 & 81.2  &  8.90 & -17.0 & 23.7 (0.1) &  6.7  (0.8) \\
HUDs-R & 219641 & 11:39:05.6 & +19:43:17 & 102.2 &  9.73 & -14.7 & 24.9 (0.3) &  3.2  (1.4) \\
HUDs-B & 219150 & 11:39:40.2 & +19:35:20 & 110.1 &  9.25 & -17.4 & 24.8 (0.1) & 10.1  (2.6) \\
HUDs-B & 215141 & 11:44:45.7 & +11:48:56 & 90.8  &  9.04 & -16.2 & 23.8 (0.1) &  4.5  (0.6) \\
HUDs-B & 219537 & 11:46:42.3 & +16:11:40 & 48.6  &  8.63 & -15.1 & 23.4 (0.1) &  4.2  (0.4) \\
HUDs-B & 219493 & 11:50:34.3 & +29:39:32 & 86.6  &  8.96 & -16.1 & 23.5 (0.1) &  3.9  (0.4) \\
HUDs-B & 219247 & 11:51:06.3 & +07:24:00 & 81.3  &  9.24 & -16.1 & 23.3 (0.1) &  3.7  (0.4) \\
HUDs-B & 219545 & 11:51:56.6 & +16:17:39 & 95.2  &  8.99 & -16.9 & 23.8 (0.1) &  5.8  (0.8) \\
HUDs-B & 219800 & 11:52:14.4 & +22:39:10 & 104.9 &  9.35 & -16.5 & 23.7 (0.1) &  4.2  (0.5) \\
HUDs-B & 215138 & 11:57:48.7 & +14:12:37 & 109.8 &  9.36 & -16.5 & 23.3 (0.1) &  3.3  (0.3) \\
HUDs-B & 215218 & 11:58:15.6 & +15:55:32 & 102.0 &  9.05 & -15.6 & 23.8 (0.1) &  3.0  (0.5) \\
HUDs-B & 215428 & 11:59:37.3 & +08:52:16 & 89.1  &  9.20 & -15.7 & 23.9 (0.1) &  3.7  (0.6) \\
HUDs-R & 229398 & 12:00:39.6 & +21:24:48 & 104.0 &  9.28 & -15.3 & 24.5 (0.1) &  3.5  (1.0) \\
HUDs-B & 229277 & 12:03:01.9 & +28:34:34 & 53.7  &  8.51 & -15.6 & 24.1 (0.1) &  6.4  (0.9) \\
HUDs-B & 229376 & 12:06:03.1 & +20:54:12 & 45.4  &  8.40 & -15.5 & 23.6 (0.1) &  5.8  (0.6) \\
HUDs-R & 229200 & 12:12:21.3 & +02:56:23 & 34.1  &  8.48 & -15.0 & 24.2 (0.1) &  8.0  (1.3) \\
HUDs-B & 229176 & 12:13:31.0 & +01:25:42 & 43.8  &  8.41 & -15.0 & 23.9 (0.1) &  5.5  (0.9) \\
HUDs-B & 223470 & 12:21:22.2 & +06:23:36 & 58.7  &  8.66 & -15.6 & 24.1 (0.1) &  5.7  (0.9) \\
HUDs-B & 223471 & 12:21:24.9 & +09:37:15 & 95.9  &  9.14 & -16.4 & 23.9 (0.1) &  4.8  (0.7) \\
HUDs-B & 227957 & 12:22:55.9 & +06:09:19 & 77.6  &  8.92 & -15.6 & 23.3 (0.1) &  3.1  (0.3) \\
HUDs-B & 223555 & 12:24:25.6 & +07:07:53 & 62.3  &  8.84 & -16.2 & 23.8 (0.1) &  6.2  (0.7) \\
HUDs-B & 227965 & 12:26:14.8 & +05:26:13 & 65.3  &  9.00 & -14.8 & 24.1 (0.1) &  3.5  (0.6) \\
HUDs-B & 223760 & 12:30:16.0 & +13:18:27 & 65.5  &  9.12 & -14.9 & 23.8 (0.1) &  3.4  (0.5) \\
HUDs-B & 226136 & 12:32:24.6 & +10:29:04 & 108.3 &  9.29 & -16.8 & 23.2 (0.1) &  3.6  (0.3) \\
HUDs-R & 749244 & 12:43:15.4 & +27:15:49 & 112.7 &  9.61 & -15.6 & 24.4 (0.1) &  3.5  (0.9) \\
HUDs-B & 749246 & 12:43:40.2 & +27:17:40 & 113.1 &  9.71 & -15.8 & 23.5 (0.1) &  2.6  (0.4) \\
HUDs-R & 229295 & 12:44:35.1 & +29:00:56 & 106.7 &  9.22 & -16.5 & 24.3 (0.1) &  5.4  (0.9) \\
HUDs-B & 229534 & 12:45:17.4 & +30:17:27 & 71.9  &  8.98 & -16.2 & 23.5 (0.1) &  4.8  (0.5) \\
HUDs-B & 229110 & 12:46:08.7 & +28:45:03 & 111.6 &  9.09 & -16.2 & 23.8 (0.1) &  3.6  (0.6) \\
HUDs-B & 227878 & 12:49:33.6 & +09:06:12 & 109.3 &  9.01 & -16.0 & 23.8 (0.1) &  3.2  (0.5) \\
HUDs-R & 223246 & 12:53:11.0 & +03:26:29 & 41.8  &  8.73 & -15.9 & 24.2 (0.1) &  9.8  (1.5) \\
HUDs-B & 226721 & 12:54:32.5 & +01:06:31 & 43.5  &  8.71 & -15.8 & 23.4 (0.1) &  6.4  (0.5) \\
HUDs-B & 233778 & 13:04:34.2 & +10:31:46 & 107.9 &  9.48 & -17.1 & 23.9 (0.1) &  6.0  (0.8) \\
HUDs-B & 231991 & 13:08:57.7 & +33:12:08 & 98.3  &  9.07 & -17.0 & 24.4 (0.1) &  7.6  (1.4) \\
HUDs-B & 239195 & 13:14:07.8 & +35:39:28 & 102.1 &  9.20 & -16.7 & 23.9 (0.1) &  5.1  (0.7) \\
HUDs-B & 238827 & 13:15:39.3 & +25:59:07 & 58.9  &  8.64 & -15.4 & 23.2 (0.1) &  3.4  (0.3) \\
HUDs-B & 239232 & 13:15:42.6 & +31:18:42 & 110.4 &  9.34 & -17.6 & 25.8 (0.2) & 16.9 (11.0) \\
HUDs-B & 238692 & 13:26:37.0 & +05:44:37 & 111.3 &  9.24 & -16.0 & 24.0 (0.1) &  3.4  (0.6) \\
HUDs-B & 238838 & 13:27:13.4 & +26:31:15 & 101.7 &  9.21 & -16.3 & 23.7 (0.1) &  4.0  (0.5) \\
HUDs-B & 239130 & 13:28:14.7 & +20:52:07 & 87.9  &  9.08 & -16.6 & 23.4 (0.1) &  4.6  (0.4) \\
HUDs-B & 239060 & 13:29:49.6 & +16:44:29 & 98.2  &  9.16 & -16.8 & 23.4 (0.1) &  4.6  (0.4) \\
HUDs-R & 233637 & 13:30:21.3 & +12:29:01 & 106.9 &  9.14 & -16.6 & 24.7 (0.1) &  6.9  (1.7) \\
HUDs-B & 239040 & 13:33:55.9 & +28:10:14 & 114.6 &  9.13 & -16.7 & 23.9 (0.1) &  4.5  (0.6) \\
HUDs-R & 238764 & 13:39:37.5 & +06:59:46 & 104.1 &  9.07 & -16.0 & 24.7 (0.1) &  5.3  (1.3) \\
HUDs-B & 239251 & 13:41:32.4 & +32:03:05 & 74.4  &  8.98 & -15.6 & 23.6 (0.1) &  3.8  (0.5) \\
HUDs-B & 238768 & 13:43:26.4 & +07:58:32 & 103.4 &  9.17 & -16.6 & 23.7 (0.1) &  4.3  (0.6) \\
HUDs-B & 238704 & 13:44:10.5 & +04:08:50 & 103.6 &  8.95 & -16.7 & 24.1 (0.1) &  5.4  (0.8) \\
HUDs-B & 238961 & 13:45:30.3 & +01:15:20 & 67.8  &  8.90 & -15.7 & 23.6 (0.1) &  4.4  (0.5) \\
HUDs-B & 233836 & 13:47:11.6 & +08:00:08 & 103.8 &  9.20 & -16.6 & 23.5 (0.1) &  4.0  (0.4) \\
HUDs-B & 239133 & 13:47:33.7 & +20:26:52 & 119.8 &  9.16 & -10.0 & 23.9 (0.1) &  5.0  (0.6) \\
HUDs-B & 239062 & 13:48:11.8 & +16:25:24 & 117.7 &  9.40 & -16.8 & 25.0 (0.2) &  7.8  (2.3) \\
HUDs-B & 238636 & 13:53:54.1 & +09:30:16 & 71.6  &  8.90 & -15.5 & 23.3 (0.1) &  3.2  (0.3) \\
HUDs-B & 238984 & 13:54:00.0 & +03:13:55 & 107.0 &  9.26 & -17.0 & 23.6 (0.1) &  4.9  (0.5) \\
HUDs-R & 232008 & 13:57:45.7 & +07:50:08 & 64.7  &  9.04 & -16.5 & 24.4 (0.1) &  9.0  (1.7) \\
HUDs-B & 243835 & 14:02:08.1 & +11:07:03 & 89.9  &  9.22 & -16.9 & 24.4 (0.1) &  8.0  (1.4) \\
HUDs-B & 248887 & 14:09:12.0 & +13:18:53 & 69.5  &  8.67 & -15.5 & 23.3 (0.1) &  3.2  (0.3) \\
HUDs-R & 749329 & 14:11:58.1 & +27:29:34 & 79.0  &  8.93 & -15.4 & 24.1 (0.1) &  4.0  (0.7) \\
HUDs-B & 249569 & 14:15:01.7 & +21:11:51 & 69.5  &  8.75 & -15.6 & 23.6 (0.1) &  4.0  (0.5) \\
HUDs-B & 749471 & 14:18:53.6 & +24:46:55 & 76.3  &  9.05 & -16.0 & 23.9 (0.1) &  5.0  (0.7) \\
HUDs-B & 248937 & 14:25:54.7 & +12:55:08 & 118.1 &  9.30 & -16.7 & 23.8 (0.1) &  4.4  (0.6) \\
HUDs-B & 245819 & 14:29:48.1 & +28:42:25 & 67.4  &  8.78 & -16.1 & 23.7 (0.1) &  5.5  (0.6) \\
HUDs-B & 249490 & 14:30:34.9 & +27:56:59 & 64.3  &  8.72 & -15.1 & 23.5 (0.1) &  3.3  (0.3) \\
HUDs-R & 249428 & 14:31:11.2 & +00:57:07 & 59.7  &  8.72 & -15.3 & 24.2 (0.1) &  5.4  (1.1) \\
HUDs-R & 242019 & 14:33:53.2 & +01:29:06 & 28.9  &  8.83 & -15.7 & 24.3 (0.1) & 13.4  (2.4) \\
HUDs-B & 248981 & 14:42:06.6 & +12:29:34 & 83.7  &  8.95 & -15.6 & 23.8 (0.1) &  3.7  (0.5) \\
HUDs-B & 248945 & 14:46:59.5 & +13:10:11 & 83.6  &  9.01 & -15.8 & 24.0 (0.1) &  4.3  (0.7) \\
HUDs-B & 249645 & 14:51:27.1 & +22:20:10 & 75.1  &  8.82 & -16.1 & 23.4 (0.1) &  4.4  (0.4) \\
HUDs-B & 749488 & 14:53:12.8 & +25:17:55 & 62.8  &  8.63 & -15.0 & 23.8 (0.1) &  3.6  (0.5) \\
HUDs-B & 249542 & 14:59:49.0 & +19:23:00 & 94.7  &  9.35 & -16.2 & 23.7 (0.1) &  4.0  (0.4) \\
HUDs-B & 749343 & 15:07:57.8 & +25:49:31 & 96.5  &  9.13 & -16.4 & 23.5 (0.1) &  4.0  (0.4) \\
HUDs-R & 258471 & 15:22:38.7 & +05:49:45 & 28.9  &  7.94 & -13.7 & 25.0 (0.2) &  7.5  (2.6) \\
HUDs-B & 252652 & 15:30:25.2 & +27:16:08 & 33.5  &  8.71 & -14.9 & 24.0 (0.1) &  7.3  (0.9) \\
HUDs-R & 257918 & 15:34:34.6 & +16:00:31 & 63.3  &  8.54 & -15.0 & 24.6 (0.1) &  5.2  (1.3) \\
HUDs-B & 253921 & 15:50:52.3 & +11:13:03 & 69.4  &  8.99 & -16.5 & 23.8 (0.1) &  6.7  (0.9) \\
HUDs-R & 253920 & 15:51:58.2 & +11:25:03 & 69.9  &  8.57 & -14.8 & 25.0 (0.2) &  5.4  (2.2) \\
HUDs-B & 749352 & 15:52:22.6 & +26:29:59 & 93.9  &  9.14 & -16.8 & 24.7 (0.1) &  8.3  (1.7) \\
HUDs-B & 258576 & 15:53:32.1 & +00:23:41 & 79.4  &  9.32 & -16.2 & 23.7 (0.1) &  4.8  (0.6) \\
HUDs-B & 258600 & 15:56:00.9 & +03:37:06 & 110.6 &  9.35 & -16.4 & 24.1 (0.1) &  4.4  (0.7) \\
HUDs-B & 262401 & 16:07:27.7 & +10:08:20 & 73.3  &  8.77 & -16.2 & 23.9 (0.1) &  5.7  (0.6) \\
HUDs-R & 268200 & 16:12:44.8 & +05:46:04 & 82.6  &  8.98 & -15.2 & 24.2 (0.2) &  3.8  (0.8) \\
HUDs-R & 749366 & 16:23:16.5 & +25:50:43 & 66.2  &  8.57 & -13.5 & 25.2 (0.3) &  3.3  (1.7) \\
HUDs-B & 749368 & 16:32:24.3 & +25:48:01 & 64.6  &  9.11 & -15.5 & 23.7 (0.1) &  4.3  (0.5) \\
HUDs-B & 310858 & 21:50:40.3 & +28:35:27 & 52.8  &  8.41 & -15.1 & 24.1 (0.1) &  5.4  (0.8) \\
HUDs-R & 312297 & 21:52:55.3 & +13:33:36 & 27.2  &  8.20 & -14.2 & 24.8 (0.1) &  9.6  (2.5) \\
HUDs-R & 748702 & 22:18:47.0 & +15:42:00 & 108.4 &  9.38 & -16.1 & 24.3 (0.1) &  4.5  (0.9) \\
HUDs-B & 323363 & 22:19:21.2 & +12:50:50 & 107.5 &  9.34 & -16.6 & 23.6 (0.1) &  4.1  (0.5) \\
HUDs-B & 323498 & 22:27:09.5 & +05:43:14 & 64.1  &  8.64 & -14.7 & 23.9 (0.1) &  3.3  (0.6) \\
HUDs-B & 321215 & 22:29:31.6 & +26:58:17 & 62.3  &  8.89 & -15.5 & 23.7 (0.1) &  4.6  (0.5) \\
HUDs-B & 322519 & 22:30:50.6 & +35:12:40 & 86.4  &  9.05 & -16.0 & 23.9 (0.1) &  4.5  (0.7) \\
HUDs-B & 323510 & 22:38:21.8 & +05:06:56 & 67.0  &  8.73 & -15.4 & 24.2 (0.1) &  4.9  (0.8) \\
HUDs-B & 323453 & 22:40:12.0 & +10:12:17 & 102.2 &  9.35 & -15.9 & 24.0 (0.1) &  3.7  (0.7) \\
HUDs-B & 322786 & 22:45:36.9 & +32:49:42 & 91.1  &  9.02 & -16.3 & 23.9 (0.1) &  4.7  (0.6) \\
HUDs-B & 322320 & 22:45:38.9 & +07:39:53 & 103.6 &  9.50 & -16.3 & 24.1 (0.1) &  4.6  (0.7) \\
HUDs-B & 321438 & 22:50:17.0 & +30:15:08 & 108.6 &  9.12 & -15.9 & 23.7 (0.1) &  3.0  (0.4) \\
HUDs-R & 321439 & 22:50:19.1 & +31:06:25 & 90.5  &  8.92 & -16.2 & 24.8 (0.1) &  6.9  (1.6) \\
HUDs-B & 322978 & 22:53:02.5 & +21:26:41 & 64.7  &  8.91 & -14.9 & 24.4 (0.2) &  4.6  (1.0) \\
HUDs-R & 321982 & 22:53:19.0 & +21:56:46 & 58.0  &  8.74 & -14.7 & 24.3 (0.1) &  4.4  (0.8) \\
HUDs-B & 321442 & 22:53:38.6 & +31:59:46 & 57.4  &  8.65 & -15.1 & 23.5 (0.1) &  3.7  (0.4) \\
HUDs-B & 322979 & 22:54:22.2 & +21:11:55 & 109.5 &  9.44 & -16.2 & 23.8 (0.1) &  3.6  (0.5) \\
HUDs-R & 322019 & 22:58:26.9 & +01:50:59 & 69.3  &  8.79 & -15.7 & 24.7 (0.1) &  6.8  (1.6) \\
HUDs-R & 322478 & 22:58:55.1 & +13:14:57 & 40.1  &  8.27 & -15.2 & 24.9 (0.1) & 10.3  (3.3) \\
HUDs-B & 334349 & 23:01:06.5 & +01:59:54 & 51.1  &  8.71 & -15.2 & 24.1 (0.1) &  5.6  (0.8) \\
HUDs-B & 334353 & 23:03:27.1 & +01:42:13 & 71.4  &  9.07 & -15.1 & 23.7 (0.1) &  3.3  (0.5) \\
HUDs-B & 333357 & 23:03:54.5 & +30:31:01 & 94.4  &  8.96 & -16.2 & 24.1 (0.1) &  4.7  (0.7) \\
HUDs-B & 748738 & 23:04:52.0 & +14:01:05 & 56.5  &  8.62 & -14.5 & 24.2 (0.1) &  3.8  (0.8) \\
HUDs-B & 334890 & 23:06:30.2 & +06:01:10 & 49.4  &  8.49 & -15.4 & 23.9 (0.1) &  5.9  (0.9) \\
HUDs-B & 336912 & 23:08:16.7 & +05:53:36 & 48.9  &  8.53 & -15.4 & 24.2 (0.1) &  6.7  (1.2) \\
HUDs-B & 748745 & 23:09:42.3 & +14:52:36 & 94.3  &  9.04 & -14.8 & 24.2 (0.2) &  2.7  (0.8) \\
HUDs-B & 333366 & 23:12:30.8 & +29:52:40 & 96.7  &  9.30 & -15.8 & 23.6 (0.1) &  3.1  (0.4) \\
HUDs-R & 336482 & 23:15:39.2 & +12:55:42 & 66.1  &  9.00 & -16.3 & 24.8 (0.1) & 10.0  (2.5) \\
HUDs-B & 748751 & 23:15:54.3 & +15:57:26 & 99.3  &  9.02 & -15.5 & 23.8 (0.1) &  2.8  (0.5) \\
HUDs-B & 333375 & 23:17:44.5 & +30:36:45 & 89.2  &  8.84 & -16.9 & 24.0 (0.1) &  6.7  (1.0) \\
HUDs-B & 332148 & 23:18:18.6 & +07:15:54 & 47.3  &  8.66 & -15.4 & 24.0 (0.1) &  6.3  (0.9) \\
HUDs-R & 334315 & 23:20:11.8 & +22:24:07 & 73.2  &  9.15 & -16.1 & 24.6 (0.1) &  7.6  (1.8) \\
HUDs-R & 332236 & 23:21:37.1 & +26:06:22 & 62.1  &  8.55 & -15.7 & 25.2 (0.2) &  9.9  (3.5) \\
HUDs-B & 748765 & 23:23:43.5 & +14:25:40 & 50.0  &  8.63 & -15.3 & 23.6 (0.1) &  4.8  (0.5) \\
HUDs-R & 336109 & 23:25:14.9 & +12:00:43 & 55.0  &  8.49 & -14.4 & 25.0 (0.2) &  5.6  (1.9) \\
HUDs-B & 748769 & 23:26:14.0 & +15:04:41 & 60.1  &  8.76 & -15.9 & 23.7 (0.2) &  5.6  (0.9) \\
HUDs-R & 337077 & 23:28:43.3 & +03:47:07 & 61.6  &  8.70 & -15.6 & 24.7 (0.2) &  7.4  (2.1) \\
HUDs-R & 336397 & 23:30:24.5 & +16:21:42 & 62.0  &  8.83 & -14.5 & 24.2 (0.1) &  3.6  (0.8) \\
HUDs-B & 333549 & 23:32:34.3 & +29:40:04 & 54.6  &  8.78 & -15.8 & 23.6 (0.1) &  5.6  (0.5) \\
HUDs-R & 336529 & 23:34:30.8 & +12:39:47 & 87.6  &  9.30 & -16.1 & 24.4 (0.1) &  5.9  (1.2) \\
HUDs-R & 333298 & 23:35:56.2 & +25:18:34 & 72.3  &  8.73 & -16.5 & 24.5 (0.1) &  8.6  (1.6) \\
HUDs-R & 335606 & 23:37:03.7 & +32:36:28 & 71.4  &  8.66 & -14.9 & 26.6 (0.6) & 11.1 (14.8) \\
HUDs-B & 333410 & 23:37:12.0 & +31:27:18 & 67.9  &  8.91 & -15.4 & 24.2 (0.1) &  5.1  (0.8) \\
HUDs-B & 334977 & 23:40:20.9 & +06:42:46 & 49.1  &  8.56 & -14.7 & 23.7 (0.1) &  3.9  (0.5) \\
HUDs-R & 336413 & 23:41:52.2 & +17:19:10 & 57.3  &  8.55 & -14.7 & 24.9 (0.2) &  5.9  (1.9) \\
HUDs-B & 333214 & 23:51:37.9 & +27:28:10 & 40.3  &  8.30 & -15.0 & 23.9 (0.1) &  5.9  (0.7) \\
HUDs-R & 332132 & 23:51:39.7 & +20:02:19 & 61.4  &  9.43 & -15.2 & 24.6 (0.2) &  5.9  (1.5) \\
HUDs-B & 333576 & 23:52:43.6 & +28:44:43 & 93.9  &  9.10 & -16.8 & 24.5 (0.1) &  7.9  (1.7) \\
HUDs-B & 336063 & 23:55:08.0 & +21:09:49 & 63.2  &  8.68 & -15.8 & 23.8 (0.1) &  5.2  (0.7) \\
HUDs-B & 332328 & 23:58:29.9 & +30:40:18 & 68.1  &  8.82 & -16.6 & 24.2 (0.1) &  8.4  (1.1) \\
\label{tab:optical}
\\ \hline
\end{supertabular}

Photometric data for galaxies in the HUDs-B and HUDs-R samples.


\bsp	
\label{lastpage}
\end{document}